\documentclass[letterpaper, paper,10pt]{AAS}		

\usepackage{bm}
\usepackage{amsmath}
\usepackage[colorlinks=true, pdfstartview=FitV, linkcolor=black, citecolor= black, urlcolor= black]{hyperref}
\usepackage{overcite}
\usepackage{footnpag}			      	

\usepackage{amsfonts}
\usepackage[utf8]{inputenc}
\usepackage{textcomp}

\usepackage{amsmath,amssymb,amsfonts}
\usepackage{bm}
\usepackage[version=4]{mhchem}
\usepackage{siunitx}
\usepackage{subcaption}
\usepackage{graphicx}
\usepackage{subcaption}
\usepackage{caption}

\usepackage{booktabs}
\usepackage{multirow}
\usepackage{longtable,tabularx}
\usepackage{adjustbox}
\usepackage{float}
\usepackage{algorithm}
\usepackage{algpseudocode}

\usepackage{tikz}
\usetikzlibrary{arrows.meta,positioning,fit,calc,shapes.geometric}

\usepackage{setspace}
\usepackage{comment}
\usepackage{footnpag}

\newcommand{\Exp}{\operatorname{Exp}}
\newcommand{\Log}{\operatorname{Log}}

\PaperNumber{26-948}

\begin{document}

\title{ Fully Multiplicative Attitude and Orbit Determination for Deep Space Navigation.}

\author{Ridma Ganganath\thanks{Graduate Research Assistant, Department of Aerospace, Iowa State University, Ames, IA 50014.}, 
\ and Simone Servadio\thanks{Assistant Professor, Department of Aerospace, Iowa State University, Ames, IA 50014.}
}

\maketitle{}



\begin{abstract}
This paper develops a geometry-consistent fully multiplicative unscented Kalman
filter (FM-UKF) for joint spacecraft attitude--orbit estimation with simultaneous
dual star-tracker misalignment calibration. The estimator uses a 21-dimensional
local error state combining attitude, angular velocity, gyroscope bias, inertial
position and velocity, and two tracker-misalignment vectors on a mixed
quaternion--Euclidean manifold. Gyroscope, star-tracker, and planet
line-of-sight measurements are fused, with celestial aberration retained to
capture velocity-dependent optical coupling. A multiplicative extended Kalman
filter (MEKF) is implemented as a first-order baseline using the same nominal
state, attitude retraction, and unit-vector measurement geometry. Monte Carlo
results show similar short-step performance, while at coarse propagation
intervals the FM-UKF remains consistent and the MEKF exhibits divergence.
\end{abstract}

\section{Introduction}
\label{sec:introduction}

Sequential Bayesian estimation is central to spacecraft guidance, navigation,
and control because it enables recursive fusion of dynamics models and noisy
measurements in real time. In spacecraft applications, this problem is naturally
geometric: attitude evolves on the rotation group, optical line-of-sight
measurements lie on the unit sphere, and orbital states, biases, and calibration
parameters evolve in Euclidean space. Classical Kalman filtering and its
nonlinear extensions provide the basic recursive framework, but their accuracy
and consistency depend strongly on how well the estimator preserves this mixed
geometry [\citenum{kalman1960,kalmanbucy1961,lefferts1982,markley2003,markley2004}].

For attitude determination, vector observations from star trackers, sun sensors,
magnetometers, and other optical instruments have long supported deterministic
and recursive estimators, including QUEST-type and quaternion-based methods
[\citenum{shuster1981,baritzhack1985,lefferts1982}]. The multiplicative
extended Kalman filter (MEKF) became a standard attitude-estimation approach
because it preserves the unit-quaternion constraint while representing
uncertainty in a minimal local error space
[\citenum{lefferts1982,markley2003,markley2004}]. However, the first-order
linearization used by the MEKF can become fragile when propagation intervals are
long, dynamics are nonlinear, or measurements couple multiple state components.
This limitation motivated unscented and higher-order nonlinear attitude
estimators, which propagate sigma points through the nonlinear model rather than
relying only on local Jacobians [\citenum{julier1997,julier2004,servadio2025quadratic,crassidis2003,crassidis2007,gui2018}].

A related issue is the geometry of the measurements. Unit-vector observations
should not be modeled as unconstrained Euclidean quantities because the measured
directions remain on \(\mathbb{S}^2\). Multiplicative measurement models and
intrinsic residuals address this by representing direction error as a small
rotation acting on the ideal line of sight, preserving unit norm and respecting
the spherical measurement geometry [\citenum{zanetti2009,zanetti2018}].
Quaternion averaging is similarly required in unscented attitude filtering so
that the propagated attitude mean is formed on the manifold rather than by
direct algebraic averaging in \(\mathbb{R}^4\) [\citenum{markley2007}].

Modern spacecraft navigation problems increasingly require estimation beyond
attitude alone. High-accuracy performance may depend on simultaneous estimation
of gyroscope bias, sensor-alignment parameters, and translational states.
Previous studies have addressed joint attitude and parameter estimation,
alignment calibration, moving-horizon estimation, and flight-calibration
methods, showing that sensor misalignment can be an observable but slowly
converging state that materially affects estimation accuracy and covariance
credibility [\citenum{vandersteen2013,smith2014,springmann2014,yoon2017}].
In deep-space optical navigation, line-of-sight observations to planets, moons,
or other celestial bodies also provide direct information about the orbital
state [\citenum{karimi2015,christian2015,andreis2024}]. In this case, the
measurement model is no longer purely an attitude problem: the inertial
reference direction depends on spacecraft position, and celestial aberration
introduces velocity-dependent coupling into the optical measurement equation
[\citenum{christian2015,andreis2024,butcher2017}].

This work develops a geometry-consistent Fully Multiplicative Unscented Kalman
Filter (FM-UKF) for joint spacecraft attitude--orbit estimation with simultaneous
calibration of two three-axis star-tracker misalignment vectors. The estimated
state contains the attitude quaternion, body angular velocity, gyroscope bias,
inertial position and velocity, and two tracker-misalignment vectors, resulting
in a 21-dimensional local error state on a mixed rotational--Euclidean manifold.
The measurement model combines two misaligned star-tracker channels with planet
line-of-sight observations, while celestial aberration is retained explicitly to
couple the optical measurements to spacecraft velocity.

The proposed FM-UKF is compared with an MEKF constructed with the same nominal
state, multiplicative attitude retraction, measurement architecture, stochastic
tuning, and intrinsic unit-vector residual
[\citenum{lefferts1982,markley2003,crassidis2003,zanetti2009,zanetti2018}].
The comparison assesses the combined effect of first-order MEKF propagation and
linearized tangent-plane measurement updates against sigma-point propagation and
multiplicative unit-vector measurement statistics in the FM-UKF. Monte Carlo
results show that both filters perform similarly for fine propagation intervals,
but differ substantially in coarse-sampling regimes. As the propagation interval
increases, the MEKF becomes overconfident and diverging, whereas the FM-UKF remains consistent
with the empirical Monte Carlo error spread. These results extend fully
multiplicative attitude-filtering ideas to a coupled
attitude--orbit--calibration problem relevant to autonomous deep-space optical
navigation [\citenum{karimi2015,christian2015,andreis2024,servadio2020recursive}].

\section{Problem Formulation}
\label{sec:problem_formulation}

The estimation problem evolves on a mixed geometric state space. The spacecraft
attitude is represented by a unit quaternion on \(\mathbb{S}^3\), optical
line-of-sight measurements lie on \(\mathbb{S}^2\), and the angular-rate, bias,
orbital, and calibration states are Euclidean. The estimated state is
\begin{equation}
    x
    \triangleq
    [q,\omega,b,r,v,\mu_1,\mu_2],
    \qquad
    q\in\mathbb{S}^3,\quad
    \omega,b,r,v,\mu_1,\mu_2\in\mathbb{R}^3 ,
    \label{eq:state_nominal}
\end{equation}
where \(q\) is the inertial-to-body attitude quaternion, \(\omega\) is the body
angular velocity, \(b\) is the gyroscope bias, \(r\) and \(v\) are the inertial
position and velocity, and \(\mu_1,\mu_2\) are the star-tracker misalignment rotation-vector coordinates. Their rotational action is
\begin{equation}
    T(\mu_j)=\exp([\mu_j]_\times)\in SO(3),
    \qquad j=1,2 .
    \label{eq:misalignment_rotation}
\end{equation}

Because the quaternion is constrained, uncertainty is represented in the
21-dimensional local error state
\begin{equation}
    e
    \triangleq
    \begin{bmatrix}
        \delta\theta^\top &
        \delta\omega^\top &
        \delta b^\top &
        \delta r^\top &
        \delta v^\top &
        \delta\mu_1^\top &
        \delta\mu_2^\top
    \end{bmatrix}^{\!\top}
    \in\mathbb{R}^{21}.
    \label{eq:error_state}
\end{equation}
For a scalar-last quaternion \(q=[q_v^\top,q_4]^\top\), the local retraction is
defined by right-multiplicative attitude injection and additive Euclidean
injection:
\begin{equation}
    x=\hat x\oplus e
    \triangleq
    \left\{
    \hat q\otimes\Exp_q(\delta\theta),\;
    \hat\omega+\delta\omega,\;
    \hat b+\delta b,\;
    \hat r+\delta r,\;
    \hat v+\delta v,\;
    \hat\mu_1+\delta\mu_1,\;
    \hat\mu_2+\delta\mu_2
    \right\},
    \label{eq:state_injection}
\end{equation}
where
\begin{equation}
    \Exp_q(\phi)
    =
    \begin{bmatrix}
        \dfrac{\sin(\|\phi\|/2)}{\|\phi\|}\phi\\[1ex]
        \cos(\|\phi\|/2)
    \end{bmatrix},
    \qquad
    \Exp_q(0)=
    \begin{bmatrix}
        0_{3\times1}\\1
    \end{bmatrix}.
    \label{eq:quat_exp}
\end{equation}
The inverse local attitude error is obtained from
\begin{equation}
    \delta q=\hat q^\ast\otimes q,
    \qquad
    \delta\theta=\Log_q(\delta q),
    \label{eq:state_difference_attitude}
\end{equation}
with the representative of \(\delta q\) chosen to have nonnegative scalar part.
The Euclidean components of \(e=x\ominus\hat x\) are obtained by ordinary subtraction. The process model combines rigid-body attitude motion, gyro-bias and
misalignment random walks, and two-body translational motion:
\begin{align}
    \dot q
    &=
    \frac{1}{2}\Omega(\omega)q,
    &
    \dot\omega
    &=
    J^{-1}\!\left(M_c-\omega\times J\omega\right),
    \label{eq:attitude_dynamics}\\
    \dot b
    &=
    w_b,
    &
    \dot\mu_1
    &=
    w_{\mu_1},
    &
    \dot\mu_2
    &=
    w_{\mu_2},
    \label{eq:calibration_dynamics}\\
    \dot r
    &=
    v,
    &
    \dot v
    &=
    -\mu_g\frac{r}{\|r\|^3}+w_a .
    \label{eq:translation_dynamics}
\end{align}
Here \(J\) is the spacecraft inertia matrix, \(M_c\) is the applied control
torque, \(\mu_g\) is the gravitational parameter, and
\(w_b,w_{\mu_1},w_{\mu_2},w_a\) are zero-mean white-noise driving processes used
in the estimator process-noise model. For the scalar-last convention,
\begin{equation}
    \Omega(\omega)=
    \begin{bmatrix}
        0          & \omega_z  & -\omega_y & \omega_x \\
        -\omega_z  & 0         & \omega_x  & \omega_y \\
        \omega_y   & -\omega_x & 0         & \omega_z \\
        -\omega_x  & -\omega_y & -\omega_z & 0
    \end{bmatrix}.
    \label{eq:omega_matrix}
\end{equation}
In the reported numerical cases, \(M_c\equiv0\), and the attitude and
translational truth trajectories are generated using torque-free rigid-body
motion and deterministic two-body orbital motion. The acceleration noise
\(w_a\) is included only in the estimator process-noise model, while the
gyroscope bias and tracker misalignment truth states are propagated as slow
random-walk calibration states.

\subsection{Gyroscope Measurement Model}
\label{subsec:gyro_measurement_model}

The gyroscope provides a direct measurement of angular velocity corrupted by bias
and white noise:
\begin{equation}
    y_g
    =
    \omega + b + \nu_g,
    \qquad
    \nu_g\sim\mathcal{N}(0,R_g).
    \label{eq:gyro_model}
\end{equation}
This measurement is processed as a Euclidean vector measurement. In contrast,
the star-tracker and planet line-of-sight measurements are processed as
unit-vector measurements on \(\mathbb{S}^2\).

\subsection{Multiplicative Direction Measurement Model}
\label{subsec:direction_measurement_model}

In this work, optical measurements are modeled as unit-vector observations corrupted by
multiplicative direction noise~[\citenum{zanetti2009,zanetti2018}]:
\begin{equation}
    y = T(\eta)h(x),
    \qquad
    h(x)\in\mathbb{S}^2,
    \qquad
    y\in\mathbb{S}^2 ,
    \label{eq:central_multiplicative_direction_model}
\end{equation}
where \(h(x)\) is the noiseless predicted line of sight and
\(T(\eta)=\exp([\eta]_\times)\) is a small rotation. This construction is
appropriate for optical direction sensors because the measurement error acts as
a pointing rotation on the unit sphere and preserves unit norm:
\begin{equation}
    \|y\|
    =
    \|T(\eta)h(x)\|
    =
    \|h(x)\|
    =
    1 .
    \label{eq:unit_norm_preservation}
\end{equation}

\begin{figure*}[t]
    \centering
    \includegraphics[width=0.8\textwidth]{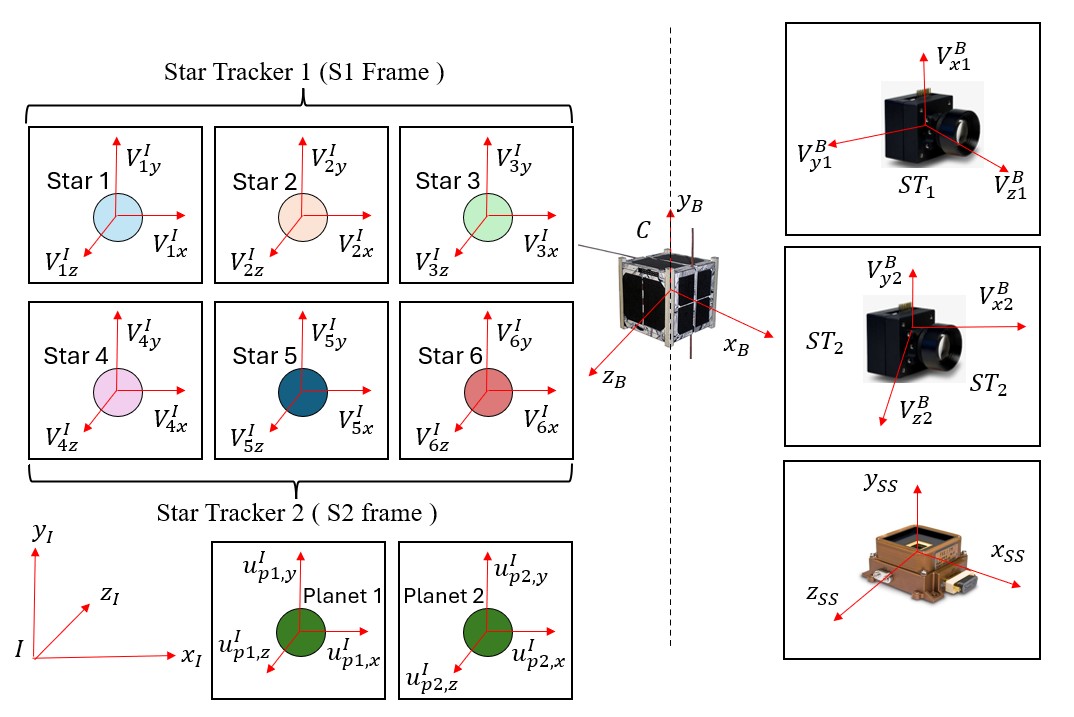}
    \caption{Measurement architecture for the joint attitude--orbit--
    misalignment estimation problem. The star-tracker measurements are affected
    by the corresponding tracker misalignment states, whereas the planet
    line-of-sight measurements are modeled as body-aligned.}
    \label{fig:measurement_architecture}
\end{figure*}

Three direction-measurement groups are considered: star tracker 1, star tracker
2, and planet line-of-sight observations. As illustrated in
Fig.~\ref{fig:measurement_architecture}, the star trackers observe known
inertial star directions through sensor frames whose orientations relative to
the spacecraft body frame are parameterized by the residual misalignment
coordinates \(\mu_1\) and \(\mu_2\). The planet line-of-sight channel provides
orbit-dependent optical measurements and is modeled as body-aligned; therefore,
the misalignment states affect only the star-tracker measurements.

For stellar observations, the inertial reference directions are assumed known.
For a planet target with known inertial position \(p_{I,m}\), the true inertial
line of sight is
\begin{equation}
    u_{I,\mathrm{true}}^{(p,m)}
    =
    \frac{p_{I,m}-r}{\|p_{I,m}-r\|},
    \qquad
    m=1,\ldots,N_p .
    \label{eq:planet_los}
\end{equation}
The apparent inertial direction after aberration is denoted by
\(u_{I,\mathrm{obs}}\). The star-tracker measurements are modeled as
\begin{align}
    y_{s_1,\ell}
    &=
    T(\eta_{s_1,\ell})
    T(\mu_1)
    T_{bi}(q)
    u_{I,\mathrm{obs}}^{(s_1,\ell)},
    \qquad
    \ell=1,\ldots,N_{s_1},
    \label{eq:star1_model}\\
    y_{s_2,\ell}
    &=
    T(\eta_{s_2,\ell})
    T(\mu_2)
    T_{bi}(q)
    u_{I,\mathrm{obs}}^{(s_2,\ell)},
    \qquad
    \ell=1,\ldots,N_{s_2}.
    \label{eq:star2_model}
\end{align}
Here \(T(\mu_j)\) maps body-frame directions into the corresponding
star-tracker sensor frame. The planet line-of-sight measurements are modeled as
\begin{equation}
    y_{p,m}
    =
    T(\eta_{p,m})
    T_{bi}(q)
    u_{I,\mathrm{obs}}^{(p,m)},
    \qquad
    m=1,\ldots,N_p .
    \label{eq:planet_model}
\end{equation}
Thus, the tracker misalignment coordinates enter only the star-tracker
channels, while the planet observations are treated as body-aligned. The direction noise is represented as a small random rotation,
\begin{equation}
    \eta
    \sim
    \mathcal{N}(0,\sigma_u^2 I_3).
    \label{eq:direction_noise}
\end{equation}
Although the rotation-noise variable is written in three dimensions, a
unit-vector measurement contains only two independent degrees of freedom because
the component of \(\eta\) parallel to the line of sight does not change the
measured direction. This rank property is handled by expressing innovations in
the tangent plane of \(\mathbb{S}^2\).

\subsection{Celestial Aberration}
\label{subsec:aberration}

Celestial aberration is retained because the apparent optical line of sight is
shifted from the true line of sight when the observer velocity is non-negligible
relative to the speed of light. Following the literature Andreis et al.[\citenum{andreis2022onboard1}], let
\(u_{\mathrm{obs}}\in\mathbb{S}^2\) denote the apparent line of sight and define
\begin{equation}
    \hat v=\frac{v}{\|v\|}.
    \label{eq:aberration_vhat}
\end{equation}
The observed angle between the apparent line of sight and the velocity direction
is
\begin{equation}
    \tan\theta_{\mathrm{obs}}
    =
    \frac{\|u_{\mathrm{obs}}\times\hat v\|}
    {u_{\mathrm{obs}}^\top\hat v}.
    \label{eq:theta_obs_tangent}
\end{equation}
The aberration angle is then computed from
\begin{equation}
    \tan\varepsilon
    =
    \frac{(\|v\|/c)\sin\theta_{\mathrm{obs}}}
    {1-(\|v\|/c)\cos\theta_{\mathrm{obs}}},
    \label{eq:aberration_tangent}
\end{equation}
where \(c\) is the speed of light. With
\(\theta_{\mathrm{true}}=\theta_{\mathrm{obs}}+\varepsilon\), the true line of
sight is recovered as
\begin{equation}
    u_{\mathrm{true}}
    =
    \frac{
    u_{\mathrm{obs}}\sin\theta_{\mathrm{true}}
    -
    \hat v\sin\varepsilon
    }
    {\sin\theta_{\mathrm{obs}}}.
    \label{eq:deaberration}
\end{equation}
Equations~\eqref{eq:theta_obs_tangent}--\eqref{eq:deaberration} define the
inverse correction
\begin{equation}
    u_{\mathrm{true}}
    =
    \mathcal{D}(u_{\mathrm{obs}},v).
    \label{eq:inverse_aberration_map}
\end{equation}

For measurement prediction, the filter requires the forward map from the true
line of sight to the apparent line of sight. This is obtained by applying the
same correction with the velocity sign reversed:
\begin{equation}
    u_{I,\mathrm{obs}}
    =
    \mathcal{A}_{\mathrm{fwd}}(u_{I,\mathrm{true}},v)
    =
    \mathcal{D}(u_{I,\mathrm{true}},-v).
    \label{eq:aberration_operator}
\end{equation}
In the numerical implementation, Eqs.~\eqref{eq:theta_obs_tangent} and
\eqref{eq:aberration_tangent} are evaluated using the equivalent
quadrant-preserving \(\operatorname{atan2}\) form. When \(\|v\|=0\), the
correction reduces to the identity; when \(\sin\theta_{\mathrm{obs}}=0\), the
limiting collinear direction is used. In the FM-UKF, sigma points are propagated
directly through Eq.~\eqref{eq:aberration_operator}; in the MEKF,
finite-difference derivatives with respect to \(u_{I,\mathrm{true}}\) and \(v\)
are used in the unit-vector measurement Jacobian.

\begin{figure}[h!]
    \centering
    \includegraphics[width=0.50\linewidth]{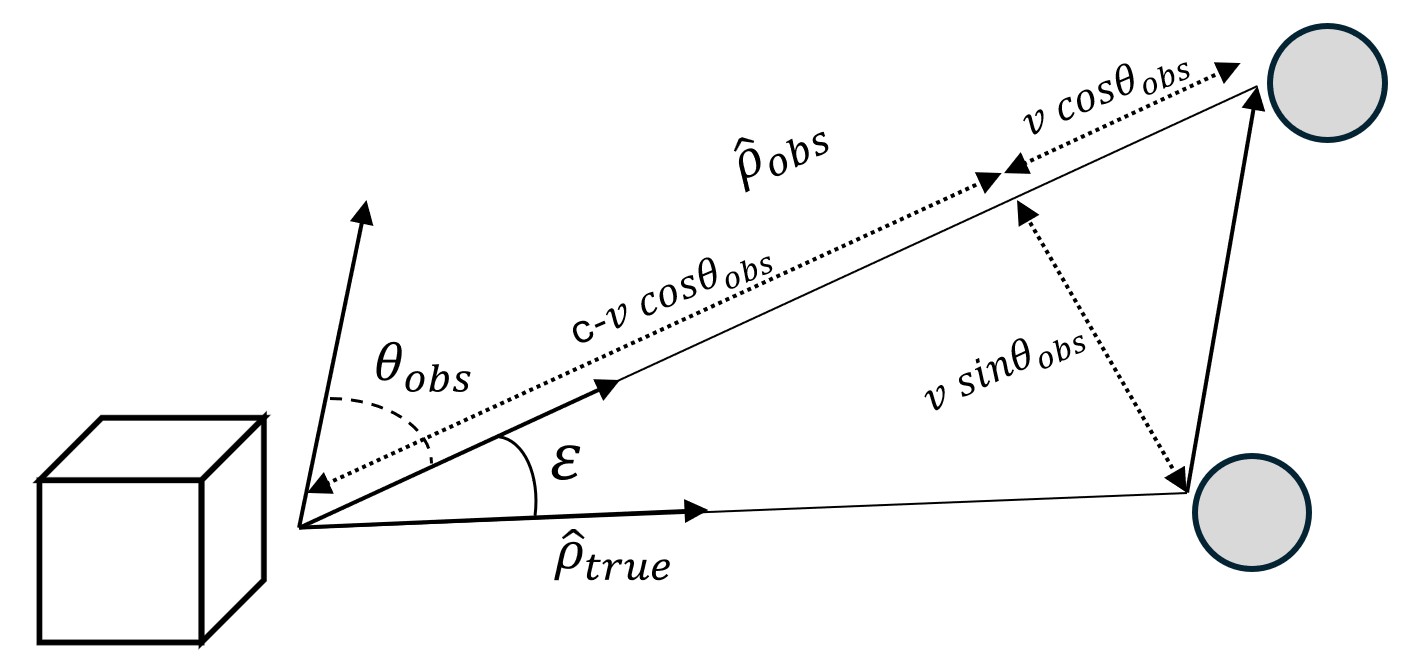}
    \caption{Geometry of celestial aberration.}
    \label{fig:aberration_geometry}
\end{figure}

\subsection{Multiplicative Residual on the Unit Sphere}
\label{subsec:intrinsic_residual}

For unit-vector measurements, innovations are formed geometrically rather than
by Euclidean subtraction. Let \(\hat u\in\mathbb{S}^2\) be the predicted
direction and \(y\in\mathbb{S}^2\) the measurement. The multiplicative residual
is
\begin{equation}
    \varepsilon_v(\hat u,y)
    \triangleq
    \frac{2(\hat u\times y)}
    {1+\hat u^\top y}.
    \label{eq:vector_residual}
\end{equation}
If \(\alpha=\cos^{-1}(\hat u^\top y)\), then
\begin{equation}
    \|\varepsilon_v(\hat u,y)\|
    =
    2\tan\left(\frac{\alpha}{2}\right).
    \label{eq:residual_norm}
\end{equation}
Thus, \(\varepsilon_v\) is a twice-Gibbs representation of the minimum
direction error. Since \(\hat u\times y\perp \hat u\), the residual has only two
independent components and lies in the tangent plane of \(\mathbb{S}^2\) at
\(\hat u\). Let \(B(\hat u)\in\mathbb{R}^{3\times2}\) satisfy
\begin{equation}
    B(\hat u)^\top B(\hat u)=I_2,
    \qquad
    B(\hat u)^\top \hat u=0 .
    \label{eq:tangent_basis}
\end{equation}
The corresponding two-dimensional tangent innovation is
\begin{equation}
    z
    =
    B(\hat u)^\top
    \varepsilon_v(\hat u,y)
    \in\mathbb{R}^2 .
    \label{eq:tangent_innovation}
\end{equation}

The MEKF uses \(z\) in Eq.~\eqref{eq:tangent_innovation} and linearizes it with
respect to the local error state. The FM-UKF instead uses the three-dimensional
multiplicative residual in Eq.~\eqref{eq:vector_residual} from transformed
unit-vector sigma points. Since these residuals have effective rank two, the
FM-UKF update uses the pseudoinverse of the residual covariance. The tangent
projection is used for the MEKF update and for diagnostic whitened innovations
and NIS calculations.

\subsection{Geometric Interpretation of Direction-Measurement Noise on
\texorpdfstring{\(\mathbb{S}^2\)}{S2}}
\label{subsec:direction_noise_geometry}

Understanding the true influence of noise on the estimation process is vital [\citenum{servadio2024likelihood}]. The residual construction in the problem formulation shows that a
unit-vector measurement has only two independent error components, both lying in
the tangent plane of \(\mathbb{S}^2\). This section uses the same tangent-plane
notation to illustrate why the multiplicative direction-noise model in
Eqs.~\eqref{eq:star1_model}--\eqref{eq:planet_model} is geometrically
preferable to additive perturbation followed by normalization.

Let \(\bar u\in\mathbb{S}^2\) be a generic nominal direction, and let
\(B(\bar u)\in\mathbb{R}^{3\times2}\) denote an orthonormal basis for
\(T_{\bar u}\mathbb{S}^2\), with the same properties as in
Eq.~\eqref{eq:tangent_basis}. Although the measurement noise can be written as
a three-dimensional small rotation, only the component orthogonal to
\(\bar u\) changes the measured direction. The component parallel to
\(\bar u\) is a rotation about the line of sight and leaves the unit vector
unchanged. Therefore, the active perturbation can be represented by the
two-dimensional tangent coordinate
\begin{equation}
    \eta \sim \mathcal{N}(0,\sigma^2 I_2),
    \label{eq:tangent_noise_source}
\end{equation}
with tangent displacement
\begin{equation}
    \zeta = B(\bar u)\eta \in T_{\bar u}\mathbb{S}^2,
    \qquad
    \varrho=\|\zeta\|=\|\eta\|.
    \label{eq:zeta_def}
\end{equation}
The equality \(\|\zeta\|=\|\eta\|\) follows from the orthonormality of
\(B(\bar u)\).

\begin{figure*}[h!]
    \centering

    \begin{subfigure}{0.80\textwidth}
        \centering
        \includegraphics[width=\linewidth]{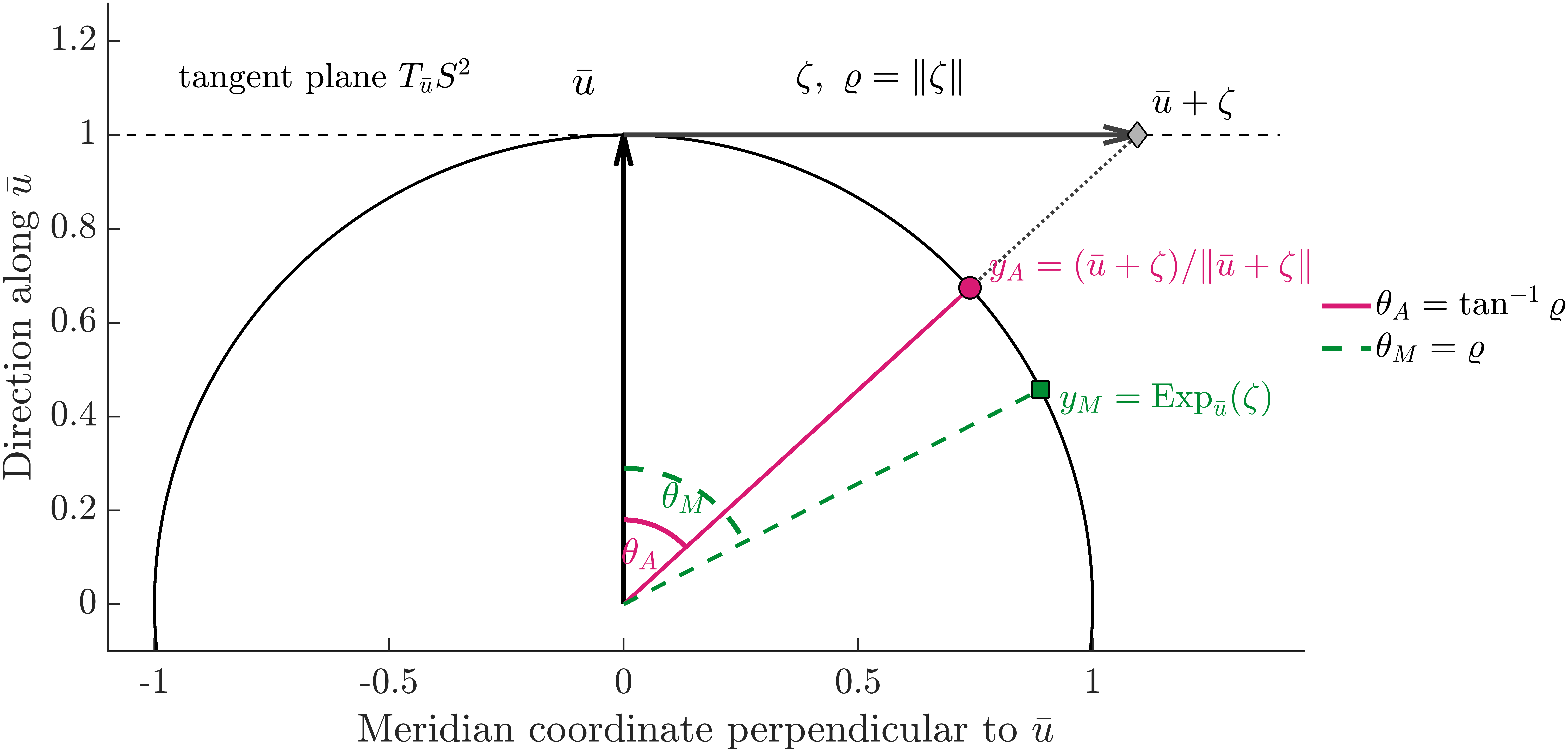}
        \caption{Meridian-plane comparison of additive and multiplicative direction-noise mappings.}
        \label{fig:mekfmukfgeo}
    \end{subfigure}

    \vspace{0.6em}

    \begin{subfigure}{0.90\textwidth}
        \centering
        \includegraphics[width=\linewidth]{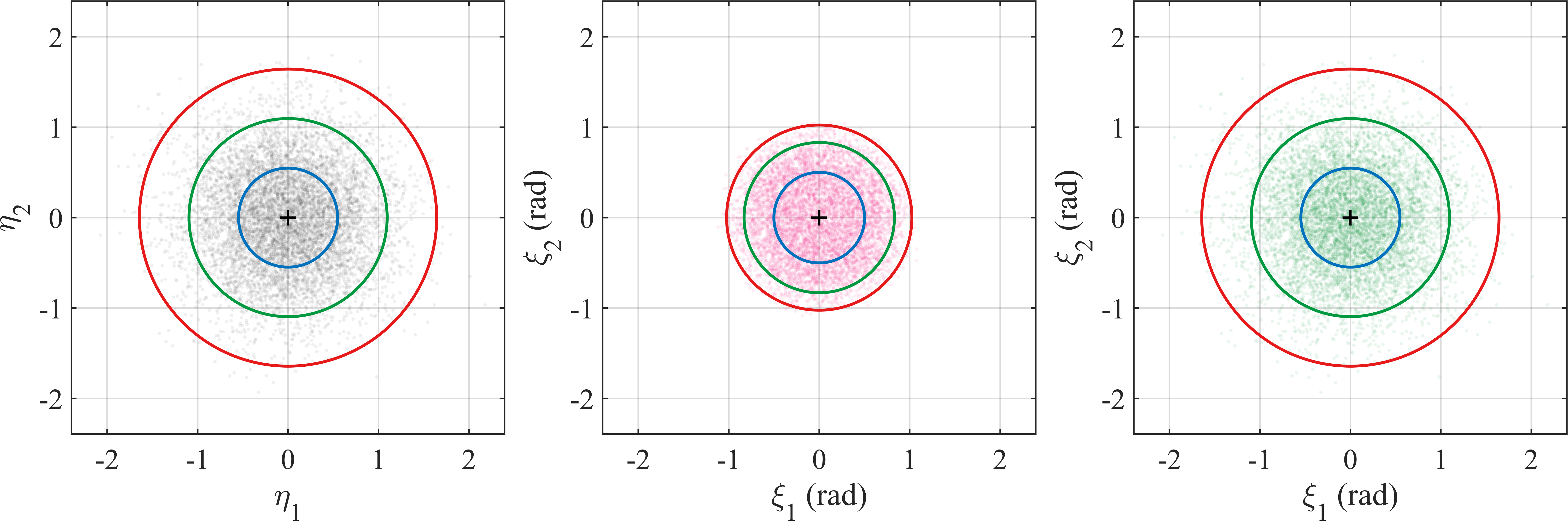}
        \caption{Tangent-plane samples and mapped directions: original samples (left), additive-normalized mapping (middle), and multiplicative mapping (right).}
        \label{fig:direction_noise_mapping_sub}
    \end{subfigure}

    \caption{Geometric interpretation of direction-measurement noise on \(\mathbb{S}^2\). }
    \label{fig:direction_noise_combined}
\end{figure*}

Let
\begin{equation}
    \theta=\cos^{-1}(\bar u^\top y)
    \label{eq:theta_def}
\end{equation}
denote the geodesic cone half-angle from \(\bar u\). Since
\(\zeta\perp\bar u\), the additive-normalized model gives
\begin{equation}
    \bar u^\top y_A
    =
    \frac{1}{\sqrt{1+\varrho^2}},
    \qquad
    \tan\theta_A=\varrho ,
    \label{eq:additive_angle_components}
\end{equation}
and therefore
\begin{equation}
    \theta_A=\tan^{-1}\varrho .
    \label{eq:theta_additive}
\end{equation}
For the multiplicative exponential-map model,
\begin{equation}
    \bar u^\top y_M=\cos\varrho ,
    \label{eq:multiplicative_angle_component}
\end{equation}
so that, on the principal geodesic branch,
\begin{equation}
    \theta_M=\varrho .
    \label{eq:theta_multiplicative}
\end{equation}
Thus, additive normalization compresses the tangent radius through
\(\theta_A=\tan^{-1}\varrho\), while the multiplicative exponential map
preserves the tangent radius as the geodesic angle. This provides the geometric motivation for using multiplicative direction noise
for unit-vector measurements. For visualization, the tangent covariance is intentionally enlarged to
\begin{equation}
    P_\eta=0.3I_2,
    \qquad
    \sigma=\sqrt{0.3}=0.5477~\mathrm{rad},
    \label{eq:visualization_sigma}
\end{equation}
so that the distortion is visible. The \(k\sigma\) source-radius
reference curves map to
\begin{align}
    \theta_A(k)&=\tan^{-1}(k\sigma),
    \label{eq:thetaA_numeric_rings}\\
    \theta_M(k)&=k\sigma,
    \qquad k=1,2,3 .
    \label{eq:thetaM_numeric_rings}
\end{align}
The additive-normalized reference radii are \(0.5011\), \(0.8309\), and
\(1.0241~\mathrm{rad}\), while the multiplicative radii are \(0.5477\),
\(1.0954\), and \(1.6432~\mathrm{rad}\). These values are used only for the geometric illustration, and the simulation sensor-noise values are specified
separately. Figure~\ref{fig:direction_noise_combined} illustrates the geometric difference
between the two direction-noise models. In Fig.~\ref{fig:direction_noise_combined}(a),
additive normalization maps the tangent perturbation back to the sphere with
\(\theta_A=\tan^{-1}\varrho\), thereby compressing the cone half-angle, whereas
the multiplicative exponential map preserves the geodesic radius,
\(\theta_M=\varrho\). Figure~\ref{fig:direction_noise_combined}(b) shows the
same effect for random tangent-plane samples: the additive-normalized mapping
contracts the local log-coordinate radii, while the multiplicative mapping
preserves the original radial structure.

\begin{figure*}[h!]
    \centering
    \includegraphics[width=0.830\textwidth]{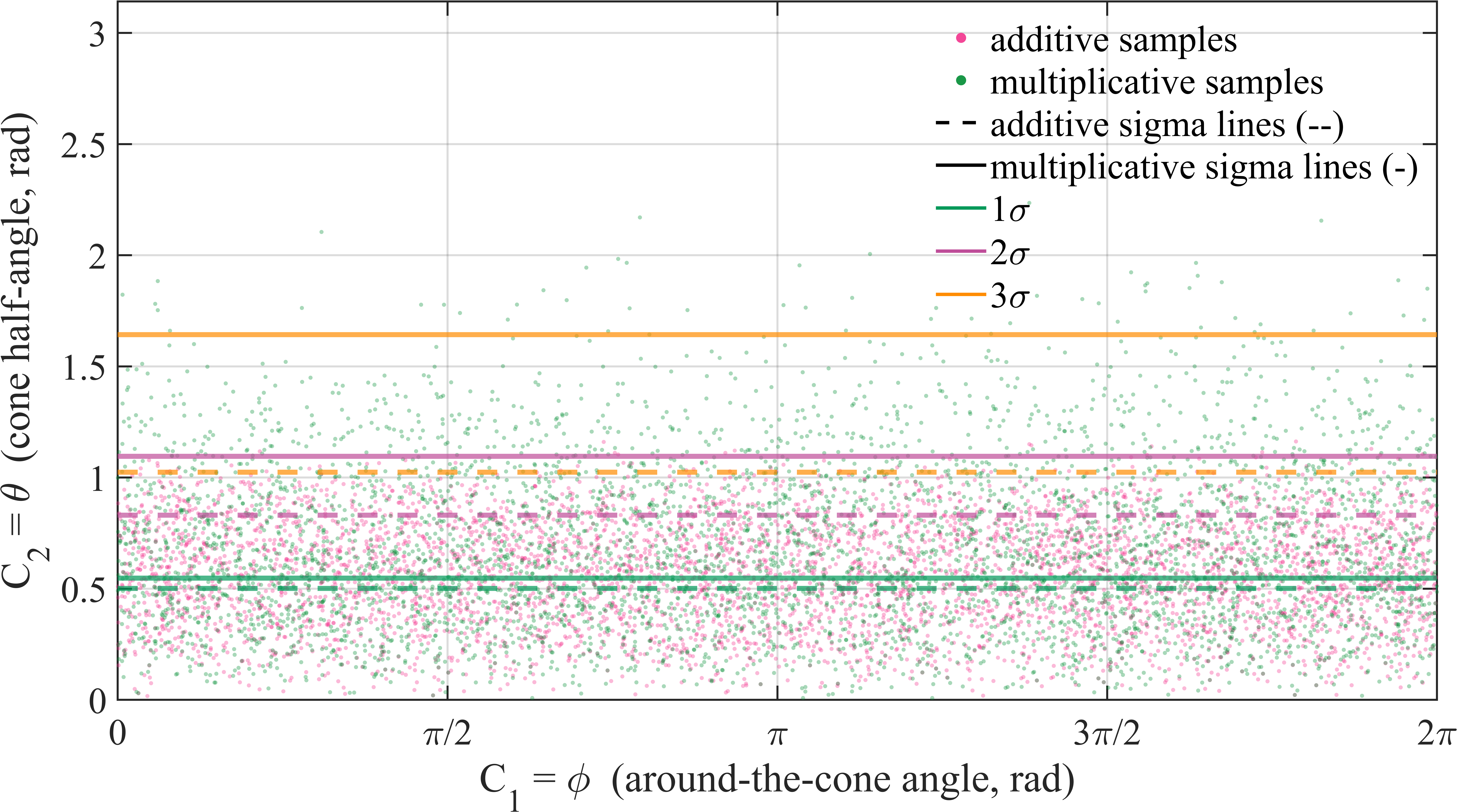}
    \caption{Unwrapped geodesic representation of the induced spherical samples.
    The coordinate \(C_1=\phi\) is the around-the-cone angle and
    \(C_2=\theta\) is the cone half-angle.}
    \label{fig:geodesic_domain_mapping}
\end{figure*}

Figure~\ref{fig:geodesic_domain_mapping} shows the same effect in geodesic polar
coordinates \((\phi,\theta)\). The additive-normalized samples shift toward
smaller cone angles, while the multiplicative samples retain the radial
structure of the original tangent perturbation. For the isotropic tangent Gaussian, the radial variable \(\varrho\) follows the
Rayleigh density
\begin{equation}
    p_\varrho(\varrho)
    =
    \frac{\varrho}{\sigma^2}
    \exp\left(-\frac{\varrho^2}{2\sigma^2}\right).
    \label{eq:rayleigh_density}
\end{equation}
Using \(\theta_M=\varrho\), the multiplicative cone-angle density is
\begin{equation}
    p_M(\theta)
    =
    \frac{\theta}{\sigma^2}
    \exp\left(-\frac{\theta^2}{2\sigma^2}\right),
    \qquad
    0\leq\theta<\pi .
    \label{eq:pm_theta}
\end{equation}
For the additive-normalized model, \(\varrho=\tan\theta\) and \(d\varrho/d\theta=\sec^2\theta\), giving
\begin{equation}
    p_A(\theta)
    =
    \frac{\tan\theta\,\sec^2\theta}{\sigma^2}
    \exp\left(-\frac{\tan^2\theta}{2\sigma^2}\right),
    \qquad
    0\leq\theta<\frac{\pi}{2}.
    \label{eq:pa_theta}
\end{equation}

\begin{figure*}[h!]
    \centering
    \includegraphics[width=0.82\textwidth]{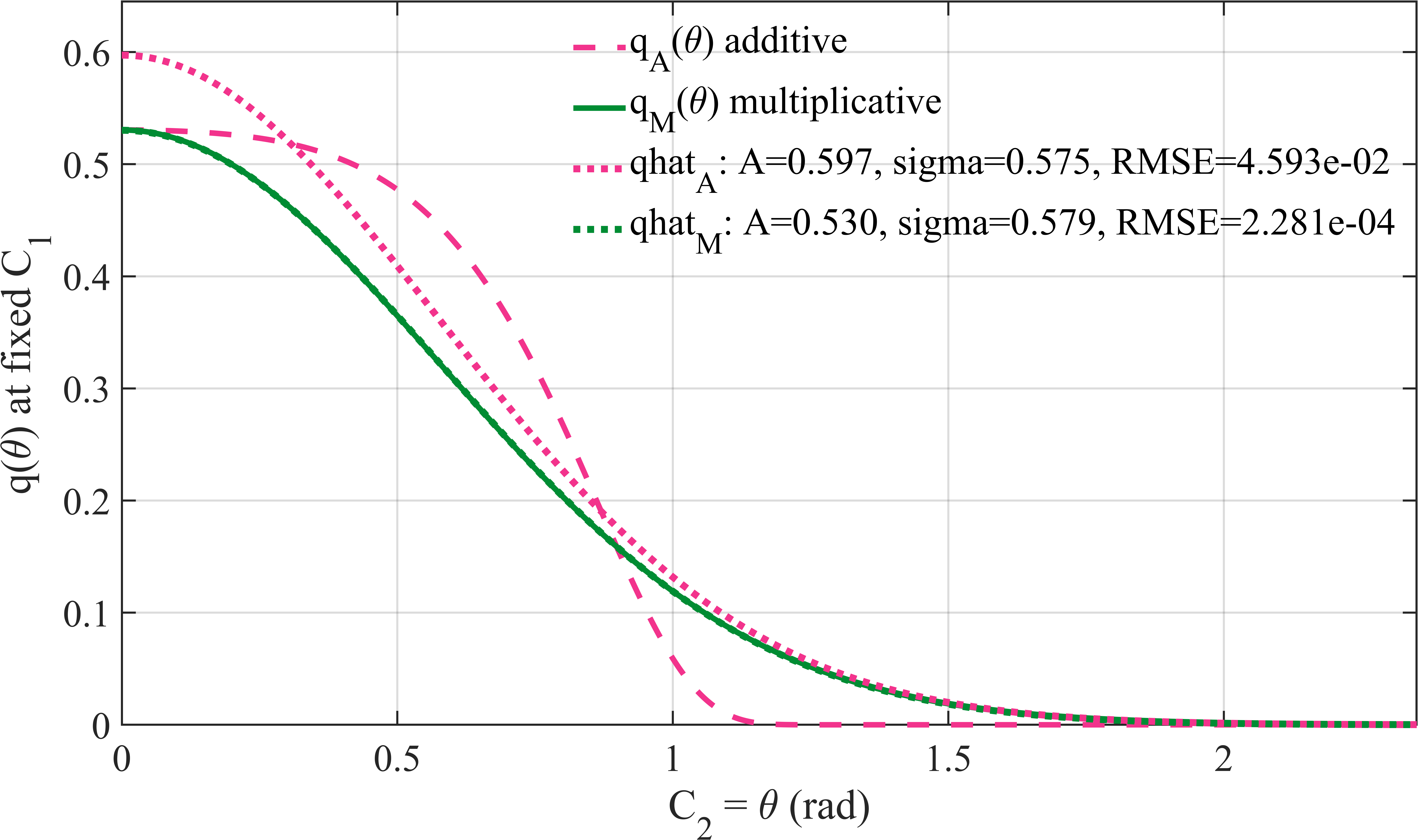}
    \caption{Pointwise densities and radial Gaussian-like fits. The
    multiplicative density is more closely matched by the chosen radial
    reference profile than the additive-normalized density.}
    \label{fig:half_gaussian_fit}
\end{figure*}

The pointwise densities on the sphere follow from the area element
\(dA=\sin\theta\,d\theta\,d\phi\). Because the distributions are axisymmetric,
\begin{equation}
    q(\theta)=\frac{p(\theta)}{2\pi\sin\theta}.
    \label{eq:pointwise_density_relation}
\end{equation}
Therefore,
\begin{equation}
    q_M(\theta)
    =
    \frac{1}{2\pi\sigma^2}
    \exp\left(-\frac{\theta^2}{2\sigma^2}\right)
    \frac{\theta}{\sin\theta},
    \qquad
    0\leq\theta<\pi ,
    \label{eq:qm_theta}
\end{equation}
and
\begin{equation}
    q_A(\theta)
    =
    \frac{1}{2\pi\sigma^2}
    \exp\left(-\frac{\tan^2\theta}{2\sigma^2}\right)
    \sec^3\theta,
    \qquad
    0\leq\theta<\frac{\pi}{2}.
    \label{eq:qa_theta}
\end{equation}
Both densities have limiting value \(1/(2\pi\sigma^2)\) at \(\theta=0\). As a visualization diagnostic, the pointwise densities are fitted with
\begin{equation}
    \hat q_H(\theta)
    =
    A\exp\!\left(-\frac{\theta^2}{2\sigma_H^2}\right),
    \qquad
    \theta\geq 0.
    \label{eq:half_gaussian_fit}
\end{equation}
The signed fitting error is
\begin{equation}
    e_q(\theta)=q(\theta)-\hat q_H(\theta),
    \label{eq:density_fit_error}
\end{equation}
and the fit quality is measured by
\begin{equation}
    \mathrm{RMSE}
    =
    \sqrt{
    \frac{1}{N}
    \sum_{i=1}^{N}
    \left(q(\theta_i)-\hat q_H(\theta_i)\right)^2 } .
    \label{eq:density_fit_rmse}
\end{equation}
This diagnostic is not used by the filter; it only visualizes the geometric
distortion caused by the two mappings. Figure~\ref{fig:half_gaussian_fit}
compares the pointwise spherical densities with radial Gaussian-like fits. For
the visualization case, the additive fit gives \(A_A=0.5973\),
\(\sigma_{H,A}=0.5751~\mathrm{rad}\), and
\(\mathrm{RMSE}_A=4.5932\times10^{-2}\), whereas the multiplicative fit gives
\(A_M=0.5302\), \(\sigma_{H,M}=0.5785~\mathrm{rad}\), and
\(\mathrm{RMSE}_M=2.2810\times10^{-4}\). Thus, over the visualized angular
interval, the multiplicative model remains much closer to the radial
Gaussian-like reference profile.

\begin{figure*}[h!]
    \centering
    \includegraphics[width=0.75\textwidth]{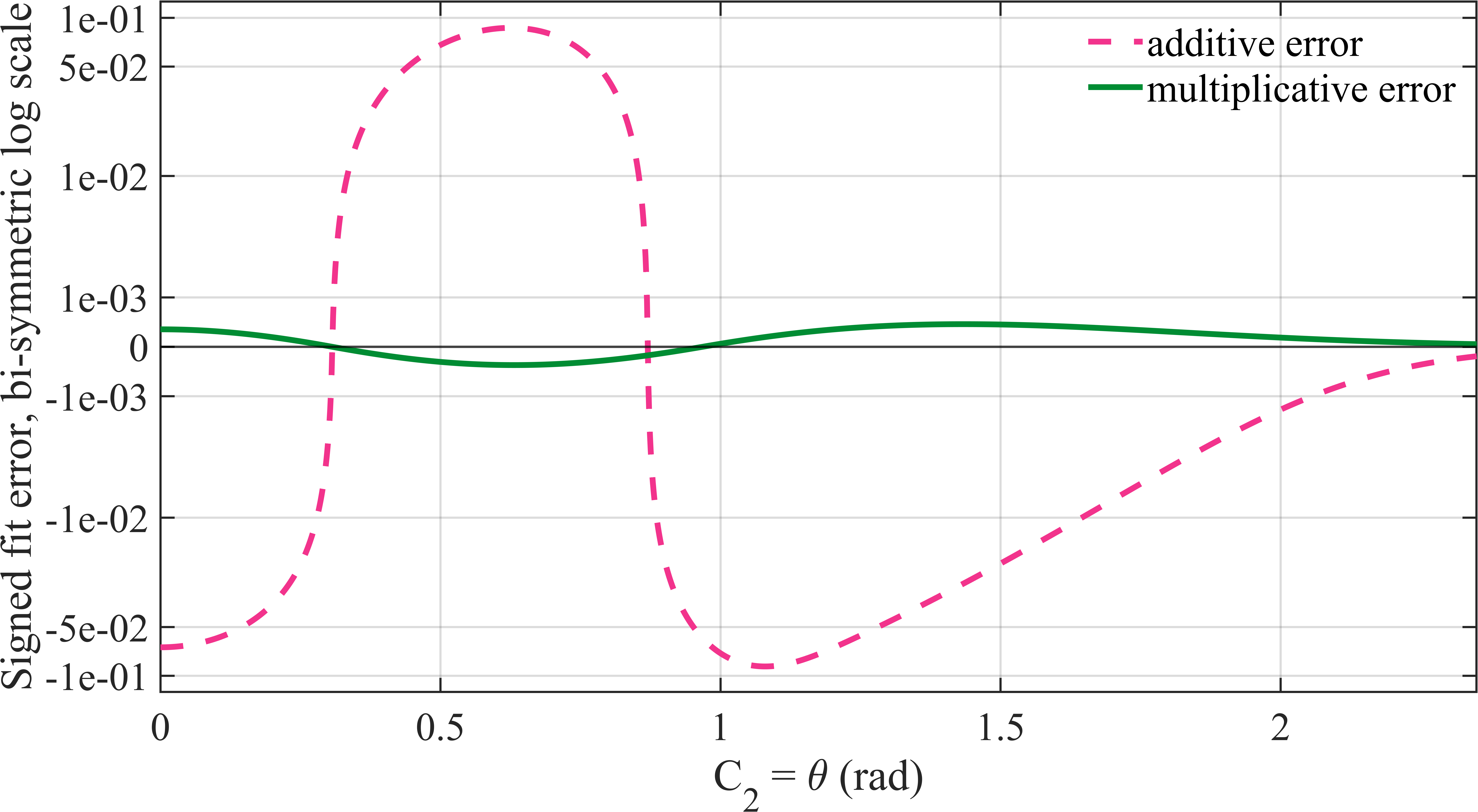}
    \caption{Signed fitting error shown on a symmetric-logarithmic scale.}
    \label{fig:half_gaussian_fit_error}
\end{figure*}

Figure~\ref{fig:half_gaussian_fit_error} shows that the additive-normalized
model produces a larger structured fitting error, while the multiplicative model
remains close to zero over the same interval. Together,
Figs.~\ref{fig:direction_noise_combined}--\ref{fig:half_gaussian_fit_error}
show that additive normalization and multiplicative rotation induce different
spherical uncertainty models. The multiplicative construction preserves the
local geodesic radius on the principal branch, whereas additive normalization
compresses it through \(\theta_A=\tan^{-1}\varrho\). This geometric distinction
motivates the multiplicative direction-noise model adopted in this work.

\section{Multiplicative Extended Kalman Filter}
\label{sec:mekf}

The multiplicative extended Kalman filter (MEKF) is used as the first-order
baseline. It uses the same nominal state, local error coordinates, and
multiplicative attitude retraction as the FM-UKF. The MEKF differs in that its
covariance prediction and unit-vector measurement updates are based on local
first-order linearization.

\subsection{Prediction}

Let \(\hat x_{k-1}^{+}\) and \(P_{k-1}^{+}\) denote the posterior nominal state
and covariance at time \(t_{k-1}\). The nominal state is propagated through the
nonlinear process model,
\begin{equation}
    \hat x_k^{-}
    =
    f_{\Delta t}\!\left(\hat x_{k-1}^{+}\right),
    \label{eq:mekf_nominal_prediction}
\end{equation}
whereas the covariance is propagated in the local error coordinates:
\begin{equation}
    \dot e
    =
    F_c e+w_c,
    \qquad
    \mathbb{E}\!\left[w_c(t)w_c(\tau)^\top\right]
    =
    Q_c\delta(t-\tau).
    \label{eq:mekf_error_dynamics}
\end{equation}
With the error-state ordering in Eq.~\eqref{eq:error_state}, the implemented
continuous-time Jacobian is
\begin{equation}
F_c =
\begin{bmatrix}
-[\hat\omega]_\times & I_3 & 0 & 0 & 0 & 0 & 0 \\
0 & A_{\omega\omega} & 0 & 0 & 0 & 0 & 0 \\
0 & 0 & 0 & 0 & 0 & 0 & 0 \\
0 & 0 & 0 & 0 & I_3 & 0 & 0 \\
0 & 0 & 0 & A_r & 0 & 0 & 0 \\
0 & 0 & 0 & 0 & 0 & 0 & 0 \\
0 & 0 & 0 & 0 & 0 & 0 & 0
\end{bmatrix},
\label{eq:mekf_Fc}
\end{equation}
where
\begin{align}
    A_{\omega\omega}
    &=
    J^{-1}\!\left([J\hat\omega]_\times-[\hat\omega]_\times J\right),
    \label{eq:mekf_Aww}\\
    A_r
    &=
    -\mu_g
    \left(
    \frac{I_3}{\|\hat r\|^3}
    -
    \frac{3\hat r\hat r^\top}{\|\hat r\|^5}
    \right).
    \label{eq:mekf_Ar}
\end{align}
The process covariance \(Q_c\) contains the angular-rate, gyro-bias,
unmodeled-acceleration, and misalignment random-walk driving-noise intensities.
The discrete transition matrix and process-noise covariance are obtained by Van
Loan discretization:
\begin{equation}
    \exp\!\left(
    \begin{bmatrix}
        F_c & Q_c\\
        0   & -F_c^\top
    \end{bmatrix}
    \Delta t
    \right)
    =
    \begin{bmatrix}
        \Phi_k & M_{12,k}\\
        0      & \Phi_k^{-\top}
    \end{bmatrix},
    \qquad
    Q_{d,k}=M_{12,k}\Phi_k^\top .
    \label{eq:mekf_vanloan}
\end{equation}
The predicted covariance is therefore
\begin{equation}
    P_k^{-}
    =
    \Phi_kP_{k-1}^{+}\Phi_k^\top+Q_{d,k}.
    \label{eq:mekf_cov_prediction}
\end{equation}

\subsection{Sequential Measurement Update}

At each epoch, the gyroscope, star-tracker, and planet line-of-sight
measurements are processed sequentially. Let \((\bar x,\bar P)\) denote the
current prior before an individual measurement update. Each MEKF update has the
standard local-error form
\begin{align}
    S_i &= H_i\bar P H_i^\top+R_i, \label{eq:mekf_generic_S}\\
    K_i &= \bar P H_i^\top S_i^{-1}, \label{eq:mekf_generic_K}\\
    \delta\hat e_i &= K_i z_i, \label{eq:mekf_generic_de}\\
    \bar x^+ &= \bar x\oplus\delta\hat e_i, \label{eq:mekf_generic_xupdate}\\
    \bar P^+
    &=
    (I-K_iH_i)\bar P(I-K_iH_i)^\top
    +
    K_iR_iK_i^\top .
    \label{eq:mekf_generic_Pupdate}
\end{align}
The posterior pair \((\bar x^+,\bar P^+)\) becomes the prior for the next
measurement at the same epoch.

For the gyroscope measurement,
\begin{equation}
    y_{g,k}
    =
    \omega_k+b_k+\nu_{g,k},
    \qquad
    \nu_{g,k}\sim\mathcal{N}(0,R_g),
    \label{eq:mekf_gyro_model}
\end{equation}
the innovation and measurement matrix are
\begin{equation}
    z_g
    =
    y_{g,k}-(\bar\omega+\bar b),
    \qquad
    H_g
    =
    \begin{bmatrix}
        0_{3\times3} & I_3 & I_3 & 0_{3\times3} &
        0_{3\times3} & 0_{3\times3} & 0_{3\times3}
    \end{bmatrix},
    \qquad
    R_i=R_g .
    \label{eq:mekf_gyro_update_compact}
\end{equation}

For a unit-vector measurement \(y_k\in\mathbb{S}^2\), the inertial reference
direction is either a fixed stellar direction or a planet line of sight. For a
planet target,
\begin{equation}
    u_I
    =
    \frac{p_I-\bar r}{\rho},
    \qquad
    \rho=\|p_I-\bar r\|,
    \qquad
    D_r
    =
    -\frac{1}{\rho}
    \left(I_3-u_Iu_I^\top\right),
    \label{eq:mekf_planet_los_compact}
\end{equation}
while \(D_r=0\) for stellar observations. Aberration is retained inside the
filter by mapping the true inertial direction to the apparent direction,
\begin{equation}
    u_{I,\mathrm{obs}}
    =
    \mathcal{A}_{\mathrm{fwd}}(u_I,\bar v).
    \label{eq:mekf_aberration_compact}
\end{equation}
The local derivatives of this mapping are evaluated numerically:
\begin{equation}
    J_u
    =
    \frac{\partial \mathcal{A}_{\mathrm{fwd}}}{\partial u_I},
    \qquad
    J_v
    =
    \frac{\partial \mathcal{A}_{\mathrm{fwd}}}{\partial v}.
    \label{eq:mekf_aberration_jacobians_compact}
\end{equation}
The body-frame direction and its position and velocity sensitivities are
\begin{equation}
    u_{B,0}=T_{bi}(\bar q)u_{I,\mathrm{obs}},
    \qquad
    H_r^B=T_{bi}(\bar q)J_uD_r,
    \qquad
    H_v^B=T_{bi}(\bar q)J_v .
    \label{eq:mekf_body_sensitivities_compact}
\end{equation}

The active channel determines the residual misalignment rotation and the planet line-of-sight channel is treated as body-aligned and therefore has
no associated misalignment state. The unit-vector innovation is formed by
projecting the intrinsic residual onto the tangent plane at \(\hat u\):
\begin{equation}
    z_k
    =
    B(\hat u)^\top
    \varepsilon_v(\hat u,y_k)
    \in\mathbb{R}^2 .
    \label{eq:mekf_unitvec_innovation}
\end{equation}
Linearizing the predicted direction with respect to the local error state gives
\begin{equation}
    \Delta\hat u
    \approx
    H_\theta^{(3)}\delta\theta
    +
    H_r^{(3)}\delta r
    +
    H_v^{(3)}\delta v
    +
    H_{\mu_1}^{(3)}\delta\mu_1
    +
    H_{\mu_2}^{(3)}\delta\mu_2 ,
    \label{eq:mekf_du}
\end{equation}
where
\begin{equation}
\begin{aligned}
H_\theta^{(3)} &= T_{\mathrm{mis}}[u_{B,0}]_\times,\\
H_r^{(3)}      &= T_{\mathrm{mis}}H_r^B,\\
H_v^{(3)}      &= T_{\mathrm{mis}}H_v^B .
\end{aligned}
\label{eq:mekf_direction_sensitivities}
\end{equation}
For an active star-tracker channel, the misalignment sensitivity is
\begin{equation}
    H_\mu^{(3)}
    =
    \left.
    \frac{\partial}{\partial\mu}
    \left(T(\mu)u_{B,0}\right)
    \right|_{\mu=\bar\mu},
    \label{eq:mekf_Hmu}
\end{equation}
which is evaluated by central differences. With the misalignment and using the tangent-plane residual linearization, the local two-dimensional
measurement model is
\begin{equation}
    z_k\approx H_ke+n_k,
    \qquad
    n_k\sim\mathcal{N}(0,R_2),
    \qquad
    R_2=\sigma_u^2I_2,
    \label{eq:mekf_unitvec_linear_model}
\end{equation}
with
\begin{equation}
H_k =
\begin{bmatrix}
J_2B^\top H_\theta^{(3)} &
0_{2\times3} &
0_{2\times3} &
J_2B^\top H_r^{(3)} &
J_2B^\top H_v^{(3)} &
J_2B^\top H_{\mu_1}^{(3)} &
J_2B^\top H_{\mu_2}^{(3)}
\end{bmatrix},
\qquad
J_2=
\begin{bmatrix}
0 & -1\\
1 & 0
\end{bmatrix},
\label{eq:mekf_Hk}
\end{equation}
where \(B=B(\hat u)\). The two zero blocks correspond to the angular-rate and
gyro-bias errors, which are not directly observed by an individual unit-vector
measurement. The matrix \(H_k\) and covariance \(R_2\) are then substituted into
Eqs.~\eqref{eq:mekf_generic_S}--\eqref{eq:mekf_generic_Pupdate}. After all
measurements at epoch \(t_k\) have been processed, the final pair is stored as
\((\hat x_k^{+},P_k^{+})\).

\section{Fully Multiplicative Unscented Kalman Filter}
\label{sec:fmukf}

The fully multiplicative unscented Kalman filter (FM-UKF) uses the same nominal
state and 21-dimensional local error coordinates as the MEKF, but constructs
prediction and measurement statistics from sigma points. The sigma points are
generated in the local error space, injected into the mixed
quaternion--Euclidean state, and propagated through the nonlinear
attitude--orbit dynamics and optical measurement models. The unit-vector update
is inspired by the fully multiplicative formulation of Zanetti and
DeMars~[\citenum{zanetti2018}] and is extended here to include orbital motion,
tracker misalignments, and aberration.

\subsection{Time Propagation}
\label{subsec:fmukf_prediction}

Let \(\hat{x}_{k-1}^{+}\) and \(P_{k-1}^{+}\) denote the posterior nominal state
and local error covariance. Process noise is included through an augmented
covariance,
\begin{equation}
    P_{\mathrm{aug},k-1}^{+}
    =
    \begin{bmatrix}
        P_{k-1}^{+} & 0\\
        0 & Q_{d,k-1}
    \end{bmatrix}.
    \label{eq:fmukf_prediction_aug_cov}
\end{equation}
Let \(\{\Xi_i^{\mathrm{aug}},w_i\}\) be the corresponding augmented sigma-point
set, with
\begin{equation}
    \Xi_i^{\mathrm{aug}}
    =
    \begin{bmatrix}
        e_i\\
        \ell_i
    \end{bmatrix},
    \qquad
    e_i,\ell_i\in\mathbb{R}^{21}.
    \label{eq:fmukf_prediction_partition}
\end{equation}
Each propagated sigma point is obtained by injecting the state perturbation,
propagating the nonlinear dynamics, and then injecting the discrete
process-noise perturbation:
\begin{equation}
    x_k^{(i)}
    =
    f_{\Delta t}
    \left(
        \hat{x}_{k-1}^{+}\oplus e_i
    \right)
    \oplus \ell_i .
    \label{eq:fmukf_prediction_sigma_propagation}
\end{equation}
Thus, the process-noise contribution is already included in the sigma-point
spread and is not added again after propagation. The predicted nominal state \(\hat{x}_k^{-}\) is obtained by averaging the
propagated sigma points on the mixed state space. The Euclidean components
\(\omega,b,r,v,\mu_1,\mu_2\) are averaged algebraically, while the quaternion is
averaged using the Markley method~[\citenum{markley2007}]:
\begin{equation}
    M_q
    =
    \sum_i w_i q_k^{(i)}q_k^{(i)\top},
    \qquad
    \hat q_k^{-}
    =
    \arg\max_{\|q\|=1} q^\top M_q q
    =
    \mathrm{eigvec}_{\max}(M_q).
    \label{eq:fmukf_markley_mean}
\end{equation}
After the predicted nominal state is formed, each propagated sigma point is
re-centered in local coordinates,
\begin{equation}
    e_k^{(i)}
    =
    x_k^{(i)}\ominus \hat{x}_k^{-},
    \label{eq:fmukf_prediction_recenter}
\end{equation}
and the predicted covariance is
\begin{equation}
    P_k^{-}
    =
    \sum_i w_i e_k^{(i)}e_k^{(i)\top}.
    \label{eq:fmukf_prediction_covariance}
\end{equation}

\subsection{Sequential Measurement Update}
\label{subsec:fmukf_sequential_update}

At each epoch, the gyroscope, star-tracker, and planet line-of-sight
measurements are processed sequentially. Let \((\bar x,\bar P)\) denote the
current prior before an individual measurement update. After each update, the
posterior pair is used as the prior for the next measurement at the same epoch.

\subsubsection{Gyroscope Update}
\label{subsubsec:fmukf_gyro_update}

For the additive gyroscope measurement,
\begin{equation}
    y_{g,k}
    =
    \omega_k+b_k+\nu_{g,k},
    \qquad
    \nu_{g,k}\sim\mathcal{N}(0,R_g),
    \label{eq:fmukf_gyro_model}
\end{equation}
sigma points \(e_i\) are generated from \(\bar P\), injected as
\(x^{(i)}=\bar x\oplus e_i\), and transformed through
\begin{equation}
    Y_g^{(i)}
    =
    \omega^{(i)}+b^{(i)} .
    \label{eq:fmukf_gyro_transformed}
\end{equation}
The predicted measurement covariance and state--measurement cross covariance are
\begin{align}
    \hat y_g
    &=
    \sum_i w_i^{(m)}Y_g^{(i)},
    \label{eq:fmukf_gyro_mean}\\
    P_{gg}
    &=
    \sum_i w_i^{(c)}
    \left(Y_g^{(i)}-\hat y_g\right)
    \left(Y_g^{(i)}-\hat y_g\right)^\top
    +R_g,
    \label{eq:fmukf_gyro_cov}\\
    P_{xg}
    &=
    \sum_i w_i^{(c)}
    e_i
    \left(Y_g^{(i)}-\hat y_g\right)^\top .
    \label{eq:fmukf_gyro_cross_cov}
\end{align}
The update is
\begin{equation}
    K_g=P_{xg}P_{gg}^{-1},
    \qquad
    \delta\hat e_g=K_g(y_{g,k}-\hat y_g),
    \qquad
    \bar x\leftarrow\bar x\oplus\delta\hat e_g,
    \qquad
    \bar P\leftarrow\bar P-K_gP_{gg}K_g^\top .
    \label{eq:fmukf_gyro_update}
\end{equation}

\subsubsection{Unit-Vector Update}
\label{subsubsec:fmukf_unit_vector_update}

For a unit-vector measurement \(y_k\in\mathbb{S}^2\), the current covariance is
augmented with multiplicative rotation-noise covariance,
\begin{equation}
    R_\eta=\sigma_u^2I_3,
    \qquad
    P_{\mathrm{aug}}
    =
    \begin{bmatrix}
        \bar P & 0\\
        0 & R_\eta
    \end{bmatrix}.
    \label{eq:fmukf_unitvec_aug_cov}
\end{equation}
Each augmented sigma point is partitioned as
\begin{equation}
    \Xi_i^{\mathrm{aug}}
    =
    \begin{bmatrix}
        e_i\\
        \eta_i
    \end{bmatrix},
    \qquad
    e_i\in\mathbb{R}^{21},
    \quad
    \eta_i\in\mathbb{R}^{3}.
    \label{eq:fmukf_unitvec_partition}
\end{equation}
The transformed direction sigma points are
\begin{equation}
    Y_i
    =
    T(\eta_i)
    h_c(\bar x\oplus e_i),
    \qquad
    Y_i\in\mathbb{S}^2 ,
    \label{eq:fmukf_transformed_direction}
\end{equation}
where \(h_c(\cdot)\) is the deterministic direction model for the active
star-tracker or planet channel, including attitude, mounting/misalignment,
position-dependent line of sight, and aberration when applicable.

The predicted direction mean \(\hat y\in\mathbb{S}^2\) is computed on the unit
sphere as the direction satisfying
\begin{equation}
    \sum_i w_i^{(m)}
    \varepsilon_v(\hat y,Y_i)
    =
    0 .
    \label{eq:fmukf_direction_mean_condition}
\end{equation}
The sigma-point residuals and measurement residual are
\begin{align}
    Z_i
    &=
    \varepsilon_v(\hat y,Y_i)
    =
    \frac{2(\hat y\times Y_i)}
    {1+\hat y^\top Y_i},
    \label{eq:fmukf_sigma_residual}\\
    z_k
    &=
    \varepsilon_v(\hat y,y_k)
    =
    \frac{2(\hat y\times y_k)}
    {1+\hat y^\top y_k}.
    \label{eq:fmukf_actual_residual}
\end{align}
Since \(Z_i\) and \(z_k\) lie in the two-dimensional tangent subspace of
\(\mathbb{S}^2\), the residual covariance is rank deficient:
\begin{align}
    P_{zz}
    &=
    \sum_i w_i^{(c)}Z_iZ_i^\top,
    \label{eq:fmukf_Pzz}\\
    P_{xz}
    &=
    \sum_i w_i^{(c)}e_iZ_i^\top .
    \label{eq:fmukf_Pxz}
\end{align}
No additive \(R_\eta\) term is added to \(P_{zz}\), because the direction noise
has already been injected through the multiplicative noise sigma points. The
gain and update are
\begin{equation}
    K_k=P_{xz}P_{zz}^{\dagger},
    \qquad
    \delta\hat e_k=K_kz_k,
    \qquad
    \bar x\leftarrow\bar x\oplus\delta\hat e_k,
    \qquad
    \bar P\leftarrow\bar P-K_kP_{zz}K_k^\top ,
    \label{eq:fmukf_unitvec_update}
\end{equation}
where \((\cdot)^\dagger\) denotes the pseudoinverse. For diagnostic consistency
checks only, \(z_k\) is projected onto a tangent basis \(B(\hat y)\) to form an
effective two-dimensional whitened innovation; this projection is not used in
the FM-UKF state update. After all measurements at epoch \(t_k\) have been processed,
\begin{equation}
    \hat x_k^{+}=\bar x,
    \qquad
    P_k^{+}=\bar P .
    \label{eq:fmukf_end_sequential_update}
\end{equation}

\section{Numerical Simulations and Results}
\label{sec:numerical_results}

The numerical study compares the baseline MEKF and the proposed FM-UKF for the
same joint attitude--orbit--calibration problem. Two propagation intervals are
considered: a short-step case with \(\Delta t=0.5~\mathrm{s}\) and a
coarse-step case with \(\Delta t=60~\mathrm{s}\), as summarized in
Table~\ref{tab:comparison_cases}. In both cases, the truth model,
initial-condition distribution, measurement architecture, and continuous-time
stochastic model are identical. The comparison therefore assesses the effect of
first-order MEKF covariance propagation relative to sigma-point propagation in
the FM-UKF. The truth model uses torque-free rigid-body attitude dynamics and two-body
orbital motion over \(T=10000~\mathrm{s}\), with spacecraft inertia
\begin{equation}
    J
    =
    \mathrm{diag}(100,60,50).
    \label{eq:J_num}
\end{equation}
At each update epoch, the measurement set consists of one gyroscope sample, two
star-tracker channels observing six inertial stars each, and three optical
planet line-of-sight observations. Celestial aberration is included in truth
measurement generation and retained inside the filter measurement model. The
gyroscope bias is modeled as a slowly varying random walk.

\begin{table}[t]
\centering
\caption{Monte Carlo comparison cases.}
\label{tab:comparison_cases}
\begin{tabular}{lccc}
\toprule
Case & \(\Delta t\) [s] & \(T\) [s] & Monte Carlo runs \\
\midrule
Short-step case  & 0.5 & 10000 & 100 \\
Coarse-step case & 60  & 10000 & 100 \\
\bottomrule
\end{tabular}
\end{table}

The tracker-misalignment states are initialized from a calibrated prior obtained
from previous MMAE-based star-tracker calibration results. Define
\begin{equation}
    \mu_{12}
    =
    \begin{bmatrix}
        \mu_1\\
        \mu_2
    \end{bmatrix},
    \qquad
    \bar\mu_{12,0}
    =
    \begin{bmatrix}
        \bar\mu_{1,0}\\
        \bar\mu_{2,0}
    \end{bmatrix}.
    \label{eq:mu12_prior_definition_num}
\end{equation}
Both the prior mean and joint covariance are imported from the MMAE-based
calibration study in prior work~[\citenum{
ganganath2025startrackermisalignmentcompensation,
ganganath2026compensatingstartrackersmisalignmentsadaptive}]. The calibrated
prior mean is
\begin{equation}
\bar\mu_{1,0}
=
\begin{bmatrix}
-0.0046336043 \\
\phantom{-}0.0021236875 \\
-0.0058639058
\end{bmatrix}
~\mathrm{rad},
\qquad
\bar\mu_{2,0}
=
\begin{bmatrix}
-0.0052031313 \\
\phantom{-}0.0020829695 \\
-0.0029448552
\end{bmatrix}
~\mathrm{rad}.
\label{eq:mu_prior_means_num}
\end{equation}
The associated calibrated covariance is denoted by
\(P_{\mu,\mathrm{prior}}\in\mathbb{R}^{6\times6}\). For each Monte Carlo run,
the true initial misalignment is sampled from the same calibrated Gaussian prior,
\begin{equation}
    \mu_{12}^{\mathrm{true}}
    \sim
    \mathcal{N}
    \left(
        \bar\mu_{12,0},
        P_{\mu,\mathrm{prior}}
    \right).
    \label{eq:mu_truth_prior_num}
\end{equation}
Thus, the previous MMAE calibration is treated as a prior calibration phase: the
filters are initialized with the calibrated mean and covariance, and the truth
misalignments are drawn consistently from that prior. The common filter initialization and stochastic tuning are summarized in
Table~\ref{tab:filter_tuning}. The initial covariance is block diagonal in the
local error coordinates, with the final \(6\times6\) block equal to
\(P_{\mu,\mathrm{prior}}\). The continuous-time process-noise model is
discretized for each propagation interval before use in both filters.

\begin{table}[t]
\centering
\caption{Common filter initialization and stochastic tuning.}
\label{tab:filter_tuning}
\begin{tabular}{lll}
\toprule
Quantity & Symbol & Value \\
\midrule
Attitude prior std. dev. & \(\sigma_{\delta\theta,0}\) & \(0.22^\circ\) per axis \\
Rate prior std. dev. & \(\sigma_{\delta\omega,0}\) & \(5\times10^{-3}~\mathrm{rad/s}\) per axis \\
Bias prior std. dev. & \(\sigma_{\delta b,0}\) & \(5\times10^{-5}~\mathrm{rad/s}\) per axis \\
Position prior std. dev. & \(\sigma_{\delta r,0}\) & \(35~\mathrm{km}\) per axis \\
Velocity prior std. dev. & \(\sigma_{\delta v,0}\) & \(2.5\times10^{-2}~\mathrm{km/s}\) per axis \\
Rate-process std. dev. & \(\sigma_{\dot\omega}\) & \(5\times10^{-7}~\mathrm{rad/s^2}\) \\
Bias-process std. dev. & \(\sigma_{\dot b}\) & \(5\times10^{-8}~\mathrm{rad/s^2}\) \\
Orbit-process std. dev. & \(\sigma_a\) & \(5\times10^{-11}~\mathrm{km/s^2}\) \\
Misalignment-process std. dev. & \(\sigma_{\dot\mu_1},\sigma_{\dot\mu_2}\) & \(10^{-14}~\mathrm{rad/s}\) \\
Gyro noise std. dev. & \(\sigma_g\) & \(10^{-4}~\mathrm{rad/s}\) \\
Star direction noise std. dev. & \(\sigma_s\) & \(5~\mathrm{arcsec}\) \\
Planet direction noise std. dev. & \(\sigma_p\) & \(10~\mathrm{arcsec}\) \\
\bottomrule
\end{tabular}
\end{table}

\subsection{Consistency, Accuracy, and Innovation Statistics}
\label{subsec:consistency_covariance}

Filter consistency is evaluated by comparing the covariance predicted by each
filter with the empirical Monte Carlo error spread [\citenum{servadio2020nonlinear,servadio2020recursive}]. For a state group
\(\mathcal{G}\in\{\delta r,\delta v,\delta\theta,\delta b,\delta\mu_1,\delta\mu_2\}\),
the predicted and empirical group standard deviations are defined as
\begin{align}
\sigma_{\mathrm{pred},\mathcal{G}}(t_k)
&=
\sqrt{
\frac{1}{N_{\mathrm{MC}}}
\sum_{m=1}^{N_{\mathrm{MC}}}
\mathrm{tr}
\left(
P_{\mathcal{G}\mathcal{G}}^{(m)}(t_k)
\right)
},
\label{eq:sigma_pred_group}\\
\sigma_{\mathrm{eff},\mathcal{G}}(t_k)
&=
\sqrt{
\sum_{j\in\mathcal{G}}
\frac{1}{N_{\mathrm{MC}}-1}
\sum_{m=1}^{N_{\mathrm{MC}}}
\left(
e_j^{(m)}(t_k)-\bar e_j(t_k)
\right)^2
}.
\label{eq:sigma_eff_group}
\end{align}
Agreement between these quantities indicates that the filter covariance is
representative of the actual Monte Carlo error spread; underprediction indicates
overconfidence.

\begin{figure*}[h!]
    \centering
    \includegraphics[width=1\textwidth]{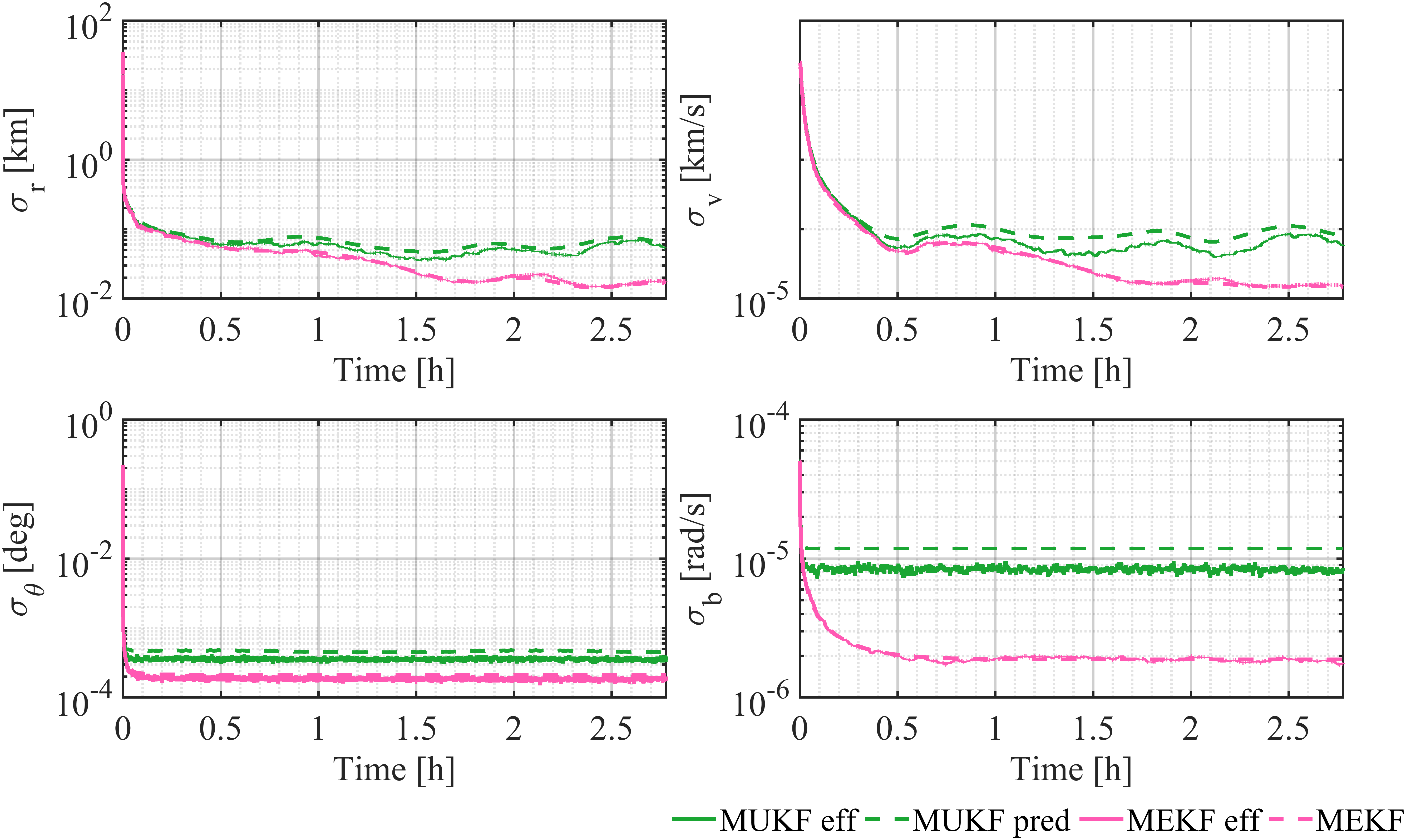}
    \caption{Predicted-versus-effective SD comparison for the short-step case
    \(\Delta t=0.5~\mathrm{s}\): position, velocity, attitude, and gyro-bias
    groups.}
    \label{fig:sd_comp_dt0.5}
\end{figure*}

\begin{figure*}[h!]
    \centering
    \includegraphics[width=0.9\textwidth]{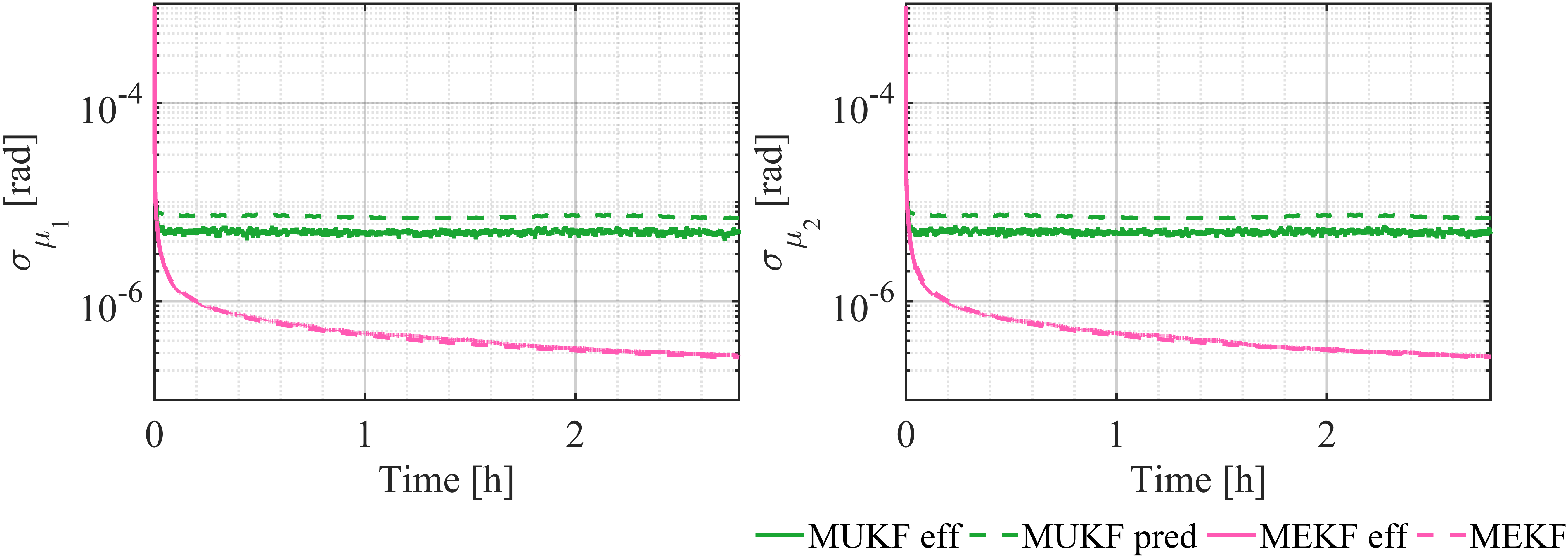}
    \caption{Predicted-versus-effective SD comparison for the short-step case
    \(\Delta t=0.5~\mathrm{s}\): tracker-misalignment groups \(\mu_1\) and
    \(\mu_2\).}
    \label{fig:sd_comp_dt0.5mu1mu2}
\end{figure*}

Figures~\ref{fig:sd_comp_dt0.5} and~\ref{fig:sd_comp_dt0.5mu1mu2} show the
short-step case, \(\Delta t=0.5~\mathrm{s}\). At this fine propagation interval,
both filters remain stable and statistically credible. The orbital errors
decrease after initialization, the attitude and gyro-bias errors remain bounded,
and the two tracker-misalignment states maintain empirical spreads close to the
reported covariance levels. Thus, for sufficiently fine propagation, the
first-order MEKF approximation is adequate and performs similarly to the FM-UKF.

\begin{figure*}[h!]
    \centering
    \includegraphics[width=1\textwidth]{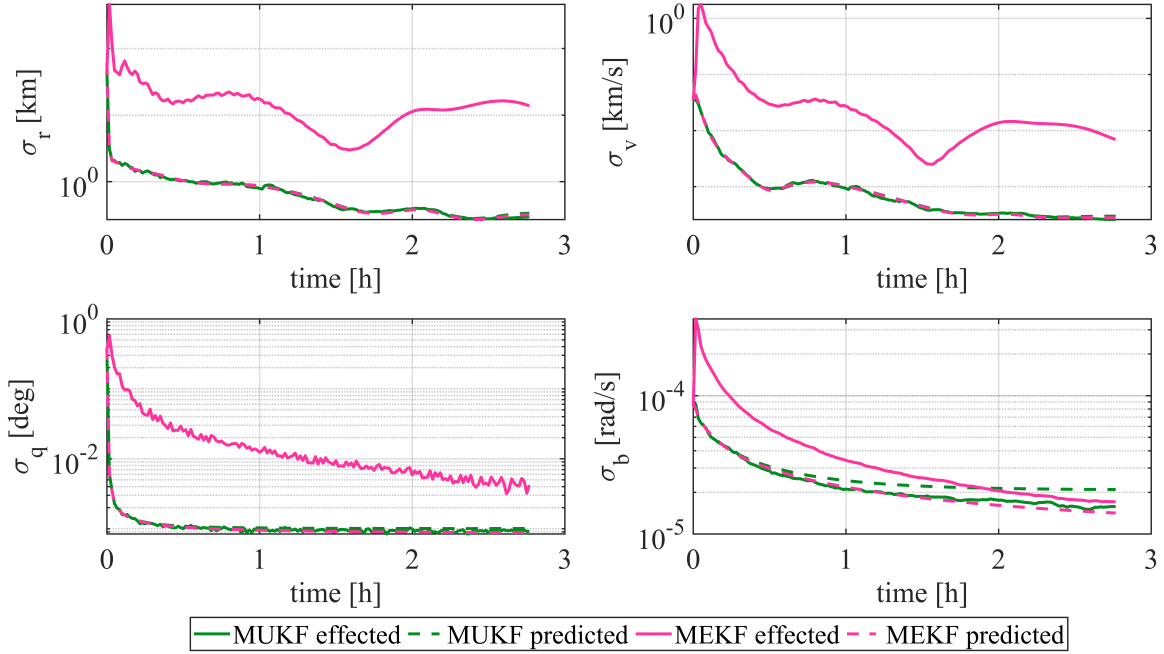}
    \caption{Predicted-versus-effective SD comparison for the coarse-step case
    \(\Delta t=60~\mathrm{s}\): position, velocity, attitude, and gyro-bias
    groups. Dashed curves denote filter-predicted standard deviations, and
    solid curves denote effective standard deviations computed from the Monte
    Carlo sample spread.}
    \label{fig:sd_comp_dt60_core}
\end{figure*}

\begin{figure*}[h!]
    \centering
    \includegraphics[width=0.9\textwidth]{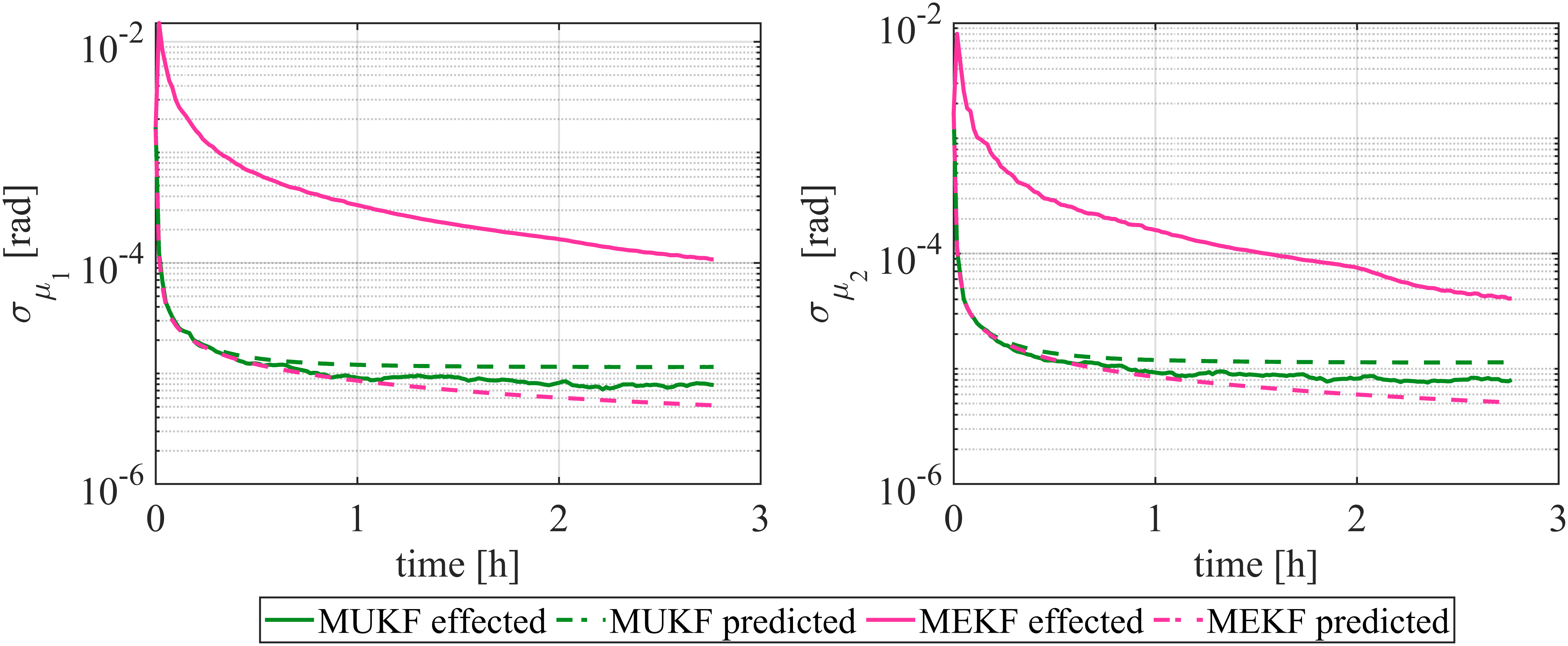}
    \caption{Predicted-versus-effective SD comparison for the coarse-step case
    \(\Delta t=60~\mathrm{s}\): tracker-misalignment groups \(\mu_1\) and
    \(\mu_2\).}
    \label{fig:sd_comp_dt60_mu}
\end{figure*}

Figures~\ref{fig:sd_comp_dt60_core} and~\ref{fig:sd_comp_dt60_mu} show the
coarse-step case, \(\Delta t=60~\mathrm{s}\). In this regime, the FM-UKF
maintains close agreement between predicted and empirical standard deviations
for the orbital, attitude, gyro-bias, and misalignment groups. The MEKF, in
contrast, becomes strongly overconfident: its empirical spread grows well above
the reported covariance, especially in the position, velocity, attitude, and
tracker-misalignment states. This behavior indicates that the first-order MEKF
propagation underestimates uncertainty accumulated over the long propagation
interval, whereas the FM-UKF preserves a more reliable nonlinear covariance
description.

Estimation accuracy is evaluated using the group RMSE,
\begin{equation}
    \Xi_{\mathcal{G}}(t_k)
    =
    \sqrt{
    \frac{1}{N_{\mathrm{MC}}}
    \sum_{m=1}^{N_{\mathrm{MC}}}
    \left\|
    e_{\mathcal{G}}^{(m)}(t_k)
    \right\|_2^2
    }.
    \label{eq:group_rmse}
\end{equation}
This metric is applied to position, velocity, local attitude error, gyro bias,
and the two tracker-misalignment states. The attitude RMSE is computed from
\(\delta\theta\) and reported in degrees.

\begin{figure*}[h!]
    \centering
    \includegraphics[width=0.9\textwidth]{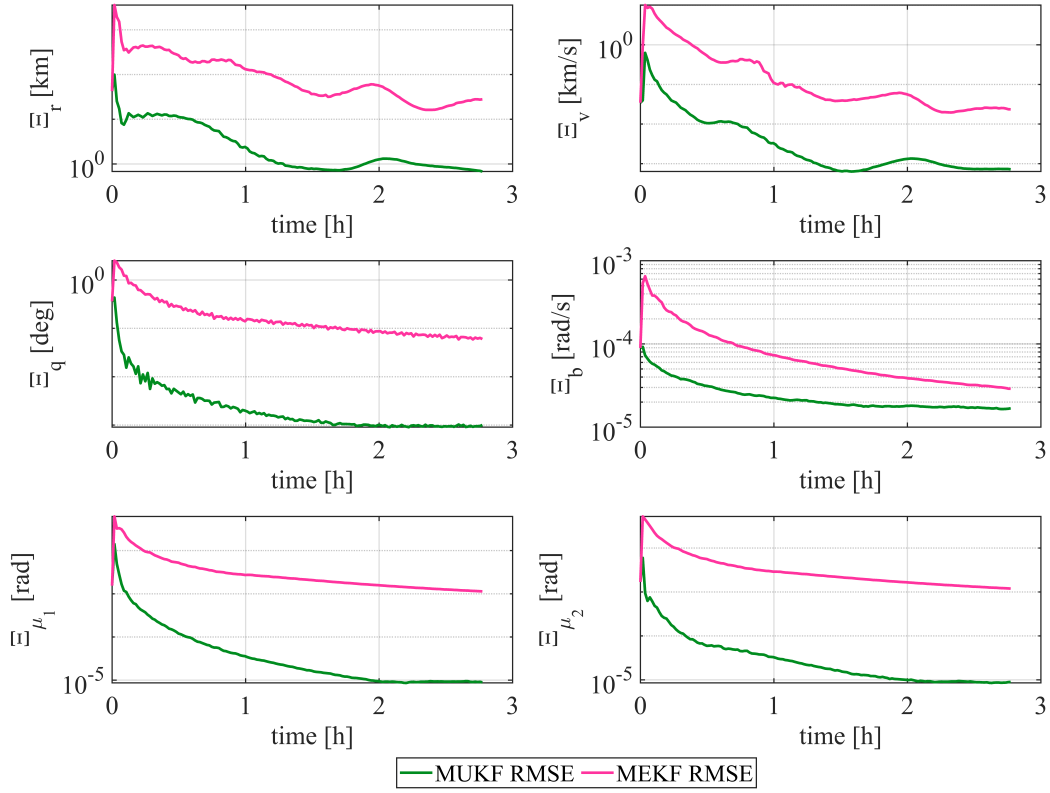}
    \caption{Group RMSE comparison for the coarse-step case
    \(\Delta t=60~\mathrm{s}\).}
    \label{fig:rmse_dt60}
\end{figure*}

Figure~\ref{fig:rmse_dt60} shows that the FM-UKF achieves lower RMSE than the
MEKF in all reported state groups for the coarse-step case. The improvement is
largest in the orbital, attitude, and tracker-misalignment states, while the
gyro-bias RMSE remains comparable because the bias is directly constrained by
the gyroscope measurement. The final RMSE values in
Table~\ref{tab:rmse_dt60} quantify this trend.

\begin{table}[h!]
\centering
\caption{Final group RMSE comparison for the coarse-step case
\(\Delta t=60~\mathrm{s}\).}
\label{tab:rmse_dt60}
\begin{tabular}{lccc}
\toprule
State group & FM-UKF RMSE & MEKF RMSE & Unit \\
\midrule
Position, \(r\)              & \(2.998\times10^{-1}\) & \(1.399\times10^{1}\)  & km \\
Velocity, \(v\)              & \(2.607\times10^{-4}\) & \(7.050\times10^{-3}\) & km/s \\
Attitude, \(\delta\theta\)   & \(9.174\times10^{-4}\) & \(4.257\times10^{-3}\) & deg \\
Gyro bias, \(b\)             & \(1.589\times10^{-5}\) & \(1.788\times10^{-5}\) & rad/s \\
Misalignment, \(\mu_1\)      & \(7.963\times10^{-6}\) & \(1.085\times10^{-4}\) & rad \\
Misalignment, \(\mu_2\)      & \(8.050\times10^{-6}\) & \(4.117\times10^{-5}\) & rad \\
\bottomrule
\end{tabular}
\end{table}

Innovation consistency is assessed using the normalized innovation squared
(NIS). At each epoch and Monte Carlo run,
\begin{equation}
    \epsilon_{\mathrm{NIS},k}^{(m)}
    =
    \nu_k^{(m)\top}
    S_k^{(m)-1}
    \nu_k^{(m)},
    \qquad
    \bar{\epsilon}_{\mathrm{NIS},k}
    =
    \frac{1}{N_{\mathrm{MC}}}
    \sum_{m=1}^{N_{\mathrm{MC}}}
    \epsilon_{\mathrm{NIS},k}^{(m)} .
    \label{eq:nis_definition}
\end{equation}
For unit-vector measurements, the diagnostic innovation is the effective
two-dimensional tangent-plane residual. For the present measurement set, the
expected value is \(n_z=33\), consisting of three gyroscope components, twelve
star-tracker unit-vector measurements with two tangent components each, and
three planet line-of-sight measurements with two tangent components each.

\begin{figure*}[h!]
    \centering
    \includegraphics[width=0.78\textwidth]{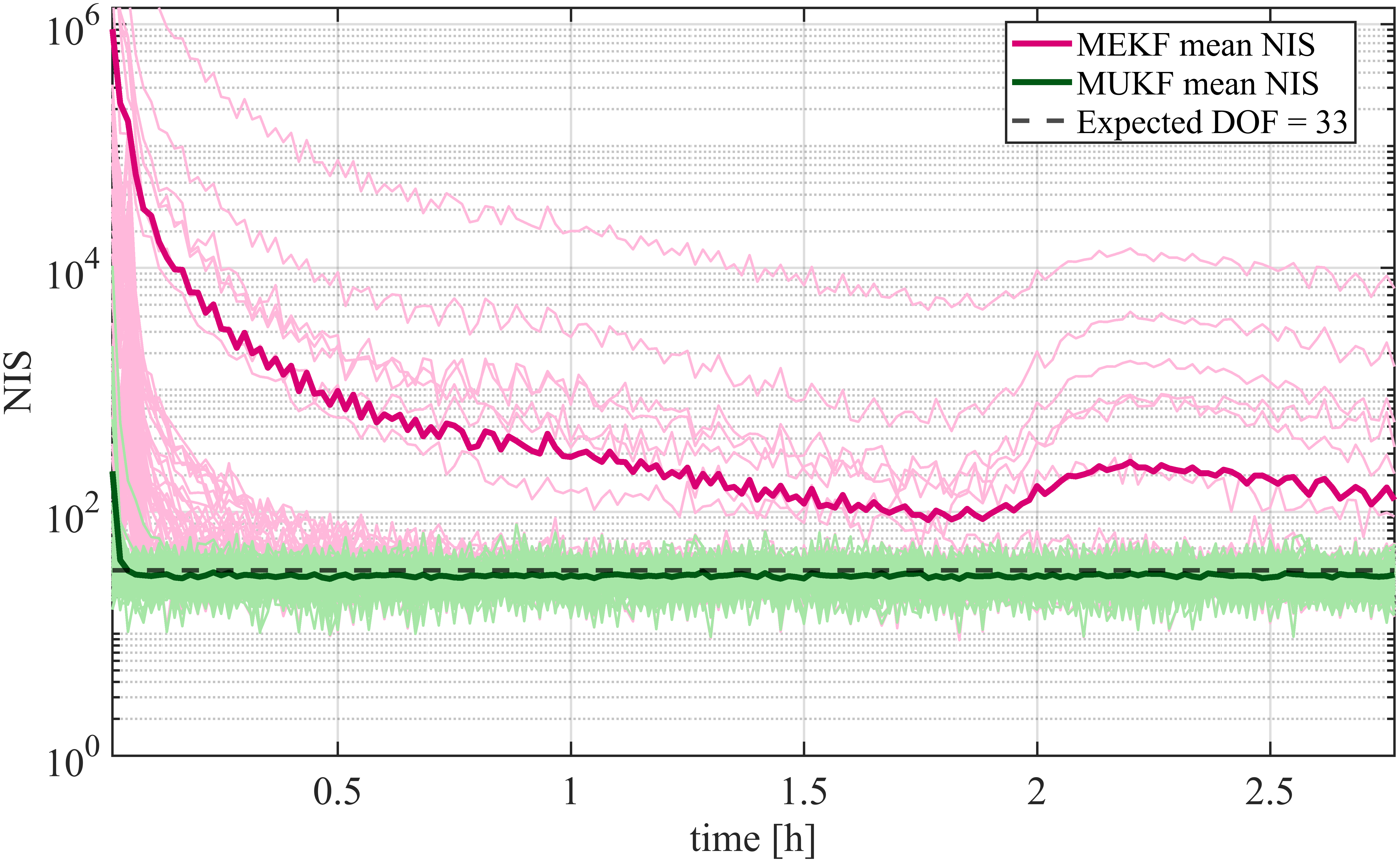}
    \caption{Mean normalized innovation squared (NIS) comparison for the
    coarse-step case \(\Delta t=60~\mathrm{s}\) over 100 Monte Carlo runs.
    Thin curves show individual Monte Carlo runs, thick curves show the Monte
    Carlo mean NIS, and the dashed black line denotes the expected measurement
    degree of freedom, \(n_z=33\).}
    \label{fig:nis_dt60}
\end{figure*}

Figure~\ref{fig:nis_dt60} confirms the covariance-consistency result. The
FM-UKF mean NIS settles near the expected value \(n_z=33\), whereas the MEKF NIS
remains orders of magnitude larger for much of the simulation. The MEKF therefore
underestimates the innovation covariance under coarse propagation, while the
FM-UKF maintains statistically credible innovations.

\begin{figure*}[h!]
    \centering
    \includegraphics[width=0.87\textwidth]{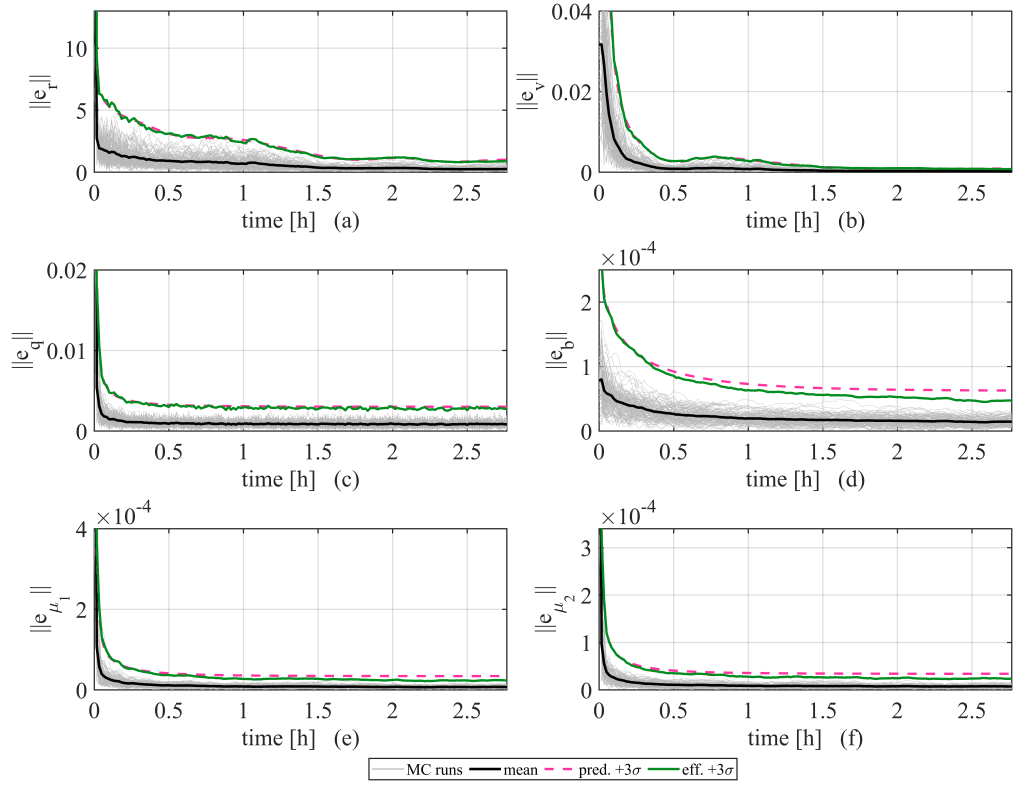}
    \caption{FM-UKF Monte Carlo norm-error convergence for the coarse-step case
    \(\Delta t=60~\mathrm{s}\). Gray curves show individual Monte Carlo runs,
    the black curve shows the ensemble mean norm error, the dashed magenta curve
    shows the filter-predicted \(3\sigma\) envelope, and the solid green curve
    shows the empirically effective \(3\sigma\) envelope.}
    \label{fig:fmukf_norm_dt60}
\end{figure*}

Finally, Fig.~\ref{fig:fmukf_norm_dt60} examines the proposed FM-UKF alone in
the coarse-step case. For each state group, the plotted Monte Carlo histories
are the norm errors
\begin{equation}
    \eta_{\mathcal{G},k}^{(m)}
    =
    \left\|
    e_{\mathcal{G},k}^{(m)}
    \right\|_2 .
    \label{eq:norm_error_history}
\end{equation}
The position and velocity errors decrease rapidly after initialization and
remain bounded, while the attitude, gyro-bias, and misalignment errors converge
to small steady-state levels. The close agreement between the predicted and
effective \(3\sigma\) envelopes further shows that the FM-UKF covariance remains
representative of the 100-run Monte Carlo spread under coarse propagation.
Together, the RMSE, covariance-spread, NIS, and norm-error results show that the
FM-UKF provides lower estimation error and more reliable uncertainty
quantification than the MEKF in the long-step regime.

\section{Conclusion}
\label{sec:conclusion}

This paper presented a geometry-consistent fully multiplicative unscented
Kalman filter for joint spacecraft attitude--orbit estimation with simultaneous
calibration of dual star-tracker misalignments. The proposed formulation
represents attitude with a unit quaternion, maintains uncertainty in a
21-dimensional local error space, models optical measurements as unit vectors on
\(\mathbb{S}^2\), and incorporates celestial aberration as a velocity-dependent
coupling in the line-of-sight measurement model. A multiplicative extended
Kalman filter was implemented as the first-order baseline, allowing the
comparison to isolate the effect of first-order covariance propagation versus
sigma-point propagation through the nonlinear attitude, orbit, calibration, and
direction-measurement models.

Monte Carlo simulations showed that the MEKF and FM-UKF perform similarly for
fine propagation intervals, where the first-order approximation remains
adequate. However, in the coarse-step case, the difference between the two
filters becomes significant. The FM-UKF preserved close agreement between
predicted and empirical covariance, maintained mean NIS values near the expected
measurement degree of freedom, and achieved lower RMSE in the orbital,
attitude, and tracker-misalignment states. In contrast, the MEKF became
systematically overconfident, with empirical error spreads and innovation
statistics that exceeded the covariance predicted by the filter. These results
demonstrate that the fully multiplicative sigma-point formulation provides a
more reliable uncertainty representation for coupled deep-space optical
navigation problems with long propagation intervals.

Future work will extend the present numerical validation toward hardware-aware
testing. The filter will be integrated with a dual-camera star/planet
image-processing pipeline and evaluated using camera-in-the-loop and
hardware-in-the-loop experiments. This next step will introduce realistic
centroiding errors, chromatic aberration, optical distortion, measurement
dropouts, and star/planet identification uncertainty. The resulting experiments
will provide a path from the present truth-model-based Monte Carlo study toward
an onboard optical navigation architecture for GPS-denied deep-space and
cislunar CubeSat missions.

\section{Acknowledgments }

The authors wish to acknowledge the support of this work through the National Aeronautics and Space Administration
(NASA) Established Program to Stimulate Competitive Research (EPSCoR) under grant number 80NSSC24M0110.

\section{Appendix}

\subsection{Calibrated Dual-Misalignment Prior Covariance}
\label{app:pmu_prior}

The calibrated joint prior covariance used for the stacked misalignment vector
\(\mu_{12}=[\mu_1^\top\ \mu_2^\top]^\top\) is
{\small
\begin{equation}
\resizebox{\textwidth}{!}{$
P_{\mu,\mathrm{prior}}=
\begin{bmatrix}
8.454257571059261\times10^{-7} & 4.668092596057423\times10^{-9} & 2.987795351006544\times10^{-8} & 5.467574073314621\times10^{-9} & -2.6238931208034446\times10^{-9} & 3.7907565804975936\times10^{-8} \\
4.668092596057423\times10^{-9} & 8.405045370489300\times10^{-7} & 1.8022770976920812\times10^{-8} & 3.296511316971029\times10^{-9} & -1.5806743552976847\times10^{-9} & 2.2866932539542806\times10^{-8} \\
2.987795351006544\times10^{-8} & 1.8022770976920812\times10^{-8} & 9.530914229287266\times10^{-7} & 2.111523143706723\times10^{-8} & -1.0130762215966536\times10^{-8} & 1.4641509798700478\times10^{-7} \\
5.467574073314621\times10^{-9} & 3.296511316971029\times10^{-9} & 2.111523143706723\times10^{-8} & 8.415563606206718\times10^{-7} & -1.8516511040202957\times10^{-9} & 2.6787611847099975\times10^{-8} \\
-2.6238931208034446\times10^{-9} & -1.5806743552976847\times10^{-9} & -1.0130762215966536\times10^{-8} & -1.8516511040202957\times10^{-9} & 8.385822108486143\times10^{-7} & -1.2855135465935025\times10^{-8} \\
3.7907565804975936\times10^{-8} & 2.2866932539542806\times10^{-8} & 1.4641509798700478\times10^{-7} & 2.6787611847099975\times10^{-8} & -1.2855135465935025\times10^{-8} & 1.0234509955701697\times10^{-6}
\end{bmatrix}.
$}
\label{eq:Pmu_prior_full}
\end{equation}

This covariance is made positive definite numerically before use and is employed both as the lower-right block of the initial filter covariance and as the covariance of the Monte Carlo truth draw for the dual-misalignment state.
\end{document}